\documentclass[twoside]{article}
\usepackage[latin1]{inputenc}
\usepackage{t1enc}
\usepackage{a4}
\usepackage{tabularx}
\usepackage{epsf}
\usepackage[]{graphicx}

\def\bref{\vspace{4pt}\noindent\hangindent=10mm}

\begin{document}

\setcounter{figure}{0}
\setcounter{section}{0}
\setcounter{equation}{0}

\begin{center}
{\Large\bf
Origins of Brown Dwarfs}\\[0.2cm]

Viki Joergens \\[0.27cm]
Sterrewacht Leiden \\
PO Box 9513, 2300 RA Leiden, The Netherlands,

and

Max-Planck-Institut f\"ur Extraterrestrische Physik \\
Giessenbachstrasse 1, 85748 Garching, Germany\\
{\tt viki@strw.leidenuniv.nl}
\end{center}

\vspace{0.5cm}

\begin{abstract}
\noindent{\it
The formation of objects below or close to the hydrogen burning 
limit is currently vividly discussed and is one of the main open 
issues in the field of the origins of stars and planets.
Applying various observational techniques, we explored a sample of 
brown dwarfs and very low-mass stars in the ChaI star forming cloud 
at an age of only a few million years and determined fundamental 
parameters for their formation and early evolution.

Tracking the question of how frequent are brown dwarf binaries and
if brown dwarfs have planets, one of the first radial velocity (RV) 
surveys of brown dwarfs sensitive down to planetary masses is carried 
out based on high-resolution spectra taken with UVES at the VLT.
The results hint at a low multiplicity fraction, which is in contrast 
to the situation for young low-mass stars.

Testing recent formation scenarios, which propose an ejection out 
of the birth place in the early accretion phase, we carried out a 
precise kinematic analysis of the brown dwarfs in our sample in 
comparison with T Tauri stars in the same field. This yielded the first empirical 
upper limit for possible ejection velocities of a homogeneous group 
of brown dwarfs.

Rotation is a fundamental parameter for objects in this early 
evolutionary phase. By means of studying the line broadening of 
spectral features in the UVES spectra as well as by tracing rotational 
modulation of their lightcurves due to surface spots in photometric
monitoring data, one of the first rotation rates of very young
brown dwarfs have been determined.

In the light of the presented observational results,
the current scenarios for the formation of brown dwarfs are
discussed
}
\end{abstract}

\section{What are brown dwarfs ?}

Brown dwarfs fill the gap between low-mass stars and giant planets in the 
mass range of about 0.08 solar masses (M$_{\odot}$) and about 13 Jupiter 
masses (M$_\mathrm{Jup}$)\footnote{1\,M$_{\odot}$=1047\,M$_\mathrm{Jup}$}, 
depending on metallicity. 
They can never fully stabilize their luminosity by hydrogen burning in 
contrast to stars and thus contract as they age until the electron gas 
in their interior is completely degenerate. At that point they have reached 
a final radius and become cooler and dimmer for their remaining life time.
In contrast to planets brown dwarfs are able to fuse deuterium, which 
defines the lower mass limit of brown dwarfs. 
It is noted that
although thermonuclear processes do not dominate the evolution of brown dwarfs, 
they do not only burn deuterium, but the more massive
ones (masses above $\sim$0.065\,M$_{\odot}$) burn lithium and may even burn
hydrogen for a while. However, they do not burn hydrogen \emph{at a rate sufficient 
to fully compensate radiative losses}.

\begin{figure}[t]
\begin{center}
\includegraphics[clip,width=0.85\textwidth]{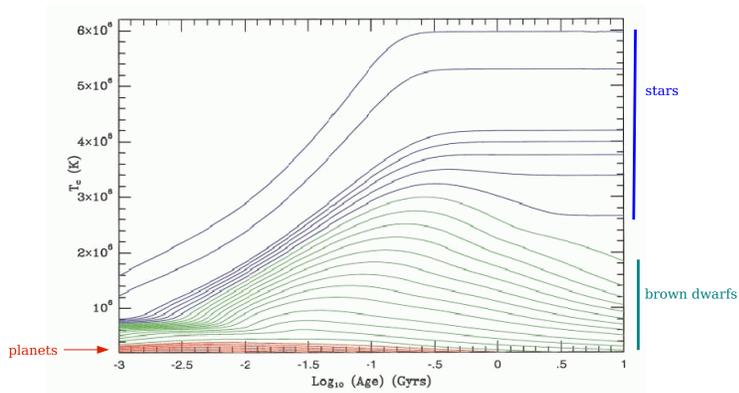}
\caption[Evolution of the central temperature]{\label{burrows_temp}
\small{{\bf Evolution of the central temperature}
for stars, brown dwarfs and planets 
from 0.3\,M$_\mathrm{Jup}$ ($\sim$ Saturn) to 0.2\,M$_{\odot}$ 
(211\,M$_\mathrm{Jup}$) versus age.
From Burrows et al. (2001).
The upper seven curves, which end in the plot with constant central temperatures,
represent stars, the 13 curves below represent brown dwarfs and the remaining
curves at the smallest
temperatures mark planets with masses $\leq$ 13\,M$_\mathrm{Jup}$.
Note the increase (classical plasma), maximum and decrease (quantum plasma) 
of the central temperature for brown dwarfs.
}}
\end{center}
\end{figure}

Fig.\,\ref{burrows_temp} shows the evolution of the central temperature 
for low-mass stars (upper seven curves with constant central temperature at 
the end of the plot), brown dwarfs and planets 
(Burrows et al. 2001).
The temperature for stars rises in the pre-main sequence phase 
due to the contraction on the Hayashi track until the 
interior is hot and dense enough to burn sufficient hydrogen to
stabilize the central temperature 
and balance gravitational pressure on the hydrogen burning main sequence.
For brown dwarfs, the central temperature also increases
due to gravitational contraction 
in the first million to a few hundred million years.
As the density in the interior increases 
part of the electron gas in the interior becomes degenerate. Electrons 
obey the \emph{Pauli exclusion principle}, allowing one 
quantum state to be occupied by only one electron. For the 
compression of partly degenerate gas, energy is \emph{needed} 
in order to bring the degenerate electrons closer. Therefore with onwardly
contraction the temperature remains constant for a while and then starts to 
decrease. 
When the electron gas is completely degenerate, no 
further contraction is possible, the brown dwarf has reached 
its final radius
and cools without compensation by compressional heating.

The existence of brown dwarfs was predicted by Kumar (1962, 1963) already in 
the early sixties. It followed a 30 year search for such faint objects
until in 1995 almost at the same time three brown dwarfs were discovered 
independently:
the Methane dwarf Gl\,229\,B, a faint companion to a nearby 
M-dwarf, as cool as 1000\,K (Nakajima et al. 1995, Oppenheimer et al. 1995) 
as well as two brown dwarfs in the Pleiades, Teide\,1 
(Rebolo et al. 1995) and PPl\,15 (Stauffer et al. 1994).
The latter object turned out to be in fact a pair of two 
gravitationally bound brown dwarfs (Basri \& Mart\'\i n 1999).
Up to now, more than 300 brown dwarfs have been detected
in star forming regions (e.g. B\'ejar et al. 1999, Comer\'on et al. 2000),
in young cluster (e.g. Pleiades, Mart\'\i n et al. 2000) and in the field
by the large infrared and optical surveys, DENIS (Delfosse et al. 1997),
2MASS (Kirkpatrick et al. 2000) and SLOAN (Hawley et al. 2002).

The year 1995 saw another spectacular discovery: the detection of the first
extrasolar planet candidate with minimum mass around $\sim$1\,M$_\mathrm{Jup}$
orbiting the sun-like star 51\,Peg (Mayor \& Queloz 1995).
Since 1995, high-precision radial velocity (RV) surveys have detected
more than 130 planets in orbit around stars 
(e.g. Mayor et al. 2003)
with masses ranging from
14 Earth masses (e.g. Santos et al. 2004) 
to the brown dwarf border at about 13\,M$_\mathrm{Jup}$. 

Before 1995, no objects were known in the mass range between
Jupiter in our own solar system and the lowest-mass stars.
Today, the situation completely changed and this
mass range is populated by hundreds of giant planets and brown dwarfs
challenging our understanding of the formation of solar systems.
Brown dwarfs play an important role in this discussion since they link the
planet population to the stellar population.

\section{Brown dwarf formation models and \\predictions}
\label{sect:formation}

The substellar border defined by the hydrogen burning mass is a crucial dividing 
line with respect to the further evolution of an object but there is no obvious reason 
why it should be of any significance for the formation mechanism by which this objects 
was produced.
Thus, by whichever process brown dwarfs are formed, it is expected to work continuously 
into the regime of very low-mass stars.
Therefore, the term \emph{formation of brown dwarfs} in this article stands for 
\emph{the formation of brown dwarfs and very low-mass stars} even if not stated explicitly.

In the traditional picture of (low-mass) star and planet formation we assume that stars 
form by collapse and hierarchical fragmentation of molecular clouds (Shu et al. 1987), 
a process which 
involves the formation of a circumstellar disk due to conservation of angular momentum.
The disk in turn is the birth place for planets, which form by condensation of dust and 
further growth by accretion of disk material. These ideas are substantially based on the 
situation in our own solar system. However, some properties of detected extrasolar planetary 
systems are difficult to explain within 
current theories of planet formation
and it is under debate if our solar system is rather the exception than the rule.
How brown dwarfs form is also an open issue but it is clear that a better knowledge 
of the mechanisms producing these transition objects between stars and planets will also
clarify some open questions in the context of star and planet formation.

Brown dwarfs may be the high-mass extension of planets
and form like giant planets in a disk around a star by either core accretion or
disk instabilities (see Wuchterl et al. 2000 for a review).
In the core accretion model, the formation of giant planets is
initiated by the condensation of solids in a circumstellar disk,
which accrete to larger bodies and form a rock/ice core.
When this core reaches the critical mass for the accretion of a gas 
envelope ($\sim$10-15\,M$_{\oplus}$) 
a so called \emph{runaway gas accretion} 
is triggered, which is only slowed down when the surrounding
gas reservoir is depleted by the accreting protoplanet.
If brown dwarfs are formed like giant planets, we should find them
in orbit around their host star. 
However, brown dwarfs are found in large number as free-floating objects.
Furthermore,
in the ongoing high-precision RV surveys for extrasolar planets,
an almost complete absence of 
brown dwarfs in close ($<$3\,AU), short-period orbits around
solar-mass stars was found
while these surveys detected more than 130 planets.
This so-called 'brown dwarf desert'
indicates that there is no continuity from planets to brown dwarfs
in terms of formation.

Brown dwarfs may also form like stars by direct gravitational collapse and fragmentation 
of molecular clouds out of cloud cores, which are cold and dense enough to become 
Jeans-unstable for brown dwarf masses. 
Such small cloud cores have not yet been detected by radio observations
but might have been missed due to insufficient detection sensitivity.
However, on theoretical grounds, 
there is a so-called opacity limit for the fragmentation, i.e. 
a limiting mass, which can become Jeans-unstable:
when reaching a certain density during the collapse the gas becomes optically thick
and is heated up leading to an increase of the Jeans-mass (Low \& Lynden-Bell 1976). 
That might prevent the formation of (lower mass) brown dwarfs by direct collapse. 
Nevertheless, assuming brown dwarfs can form in that way, 
they would be the low-mass extension of the stellar population
and should, therefore, have similar (scaled down) properties as young low-mass stars
(T~Tauri stars). 
We know that T~Tauri stars have a very high multiplicity fraction, in some star forming 
regions close to 100\% (e.g. Leinert et al. 1993, Ghez et al. 1993, 1997,
K\"ohler et al. 2000) indicating that the vast majority of low-mass stars is 
formed in binaries or higher order multiple systems.
The stellar companions are thought to be formed by disk fragmentation or
filament fragmentation in the circumstellar disk created by the collapse
of the molecular cloud core.
Therefore, we would also expect a high multiplicity fraction of brown dwarfs
in this formation model.
Furthermore, brown dwarfs should have 
circumstellar disks due to preservation of angular momentum during the collapse,
harbour planets, 
and should have the same kinematical properties
as the T~Tauri stars in the same field.

Another idea is that brown dwarfs formed by direct collapse of 
unstable cloud cores of stellar masses and would 
have become stars if the accretion process was not stopped at an early 
stage by an external process before the object has accreted to stellar mass.
It was proposed that such an external process can be the ejection of the 
protostar out of the dense gaseous environment due to dynamical interactions
(Reipurth \& Clarke 2001).
It is known that the dynamical evolution of gravitationally 
interacting systems of three or more bodies leads to frequent close two-body 
encounters and to the formation of close binary pairs 
out of the most massive objects in the system 
as well as to the ejection of the lighter bodies into extended orbits or out of 
the system with escape velocity (e.g. Valtonen \& Mikkola 1991).
The escape of the lightest body is an expected outcome since the
escape probability scales approximately as the inverse third power of the mass.
Sterzik \& Durisen (1995, 1998) and Durisen et al. (2001)
considered the formation of run-away T~Tauri stars
by such dynamical interactions of compact clusters. 

The general expectations for the properties of brown dwarfs formed by the ejection 
scenario are a low binary frequency (maybe 5\%), no wide brown dwarf binaries 
and only close-in disks ($<$ 5-10\,AU)
since companions or disk material at larger separations will be truncated by the 
ejection process.
Furthermore, the kinematics of ejected brown dwarfs 
might differ from non-ejected members of the cluster. 
Hydrodynamical calculations have shown that the collapse of a molecular
cloud can produce brown dwarfs in this way
(Bate et al. 2003, Delgado-Donate et al. 2003, Bate \& Bonnell 2005), while
N-body simulations of the dynamical decay allowed
the prediction of statistically significant properties of ejected brown dwarfs
(Sterzik \& Durisen 2003, Delgado-Donate et al. 2004,
Umbreit et al. 2005).
However, the predictions differ significantly among the different models;
this is further discussed in Sect.\,\ref{sect:kinematic}.

An external process, which prevents the stellar embryo from further 
growth in mass can also be a strong UV wind from a nearby 
hot O or B star, which ionizes and photoevaporates the surrounding gas
(Kroupa \& Bouvier 2003, Whitworth \& Zinnecker 2004).
A significant disruption of the accretion envelope by photoevaporation
will also lead generally to a low multiplicity fraction and
limited disk masses. 
Since there is no such a hot star in the Cha\,I cloud, 
the brown dwarfs in this region cannot have been formed
by this mechanism.

The various ideas for the formation of brown dwarfs need to be constrained by
observations of brown dwarfs, key parameters are among others
the multiplicity and kinematics, which have been studied for young brown dwarfs in
Cha\,I as described in the following.  
Furthermore, rotation is an important parameter for the early evolution and it might
reflect the interaction with a disk due to magnetic disk braking.

\section{Comprehensive observations of brown dwarfs and
very low-mass stars in Cha\,I}

\subsection{Sample}

The observation of very young brown dwarfs and very low-mass stars
allows insights into the formation and early evolution below or close to the
substellar limit.
One of the best grounds for such a study 
is the Cha\,I cloud, which is part of the larger Chamaeleon complex.
At a distance of 160\,pc, it is one of the closest sites of active
low-mass star and brown dwarf formation.
Comer\'on and coworkers initiated here one of the first surveys for young very 
low-mass objects down to the substellar regime by means of an H$\alpha$ objective 
prism survey (Comer\'on et al. 1999, 2000; Neuh\"auser \& Comer\'on 1998, 1999).
They found twelve very low-mass M6--M8--type objects,
Cha\,H$\alpha$\,1 to 12, in the center of Cha\,I with ages of 1 to 5\,Myrs,
among which are four bona fide brown dwarfs and six brown dwarf candidates. 
Furthermore, they found or confirmed several very low-mass stars with masses smaller 
than 0.2\,M$_{\odot}$ and spectral types M4.5--M5.5 in Cha\,I 
(B\,34, CHXR\,74, CHXR73 and CHXR\,78C) as well as one 0.3 solar mass M2.5 star (Sz\,23).

The membership to the Cha\,I cloud and therefore the youth of the objects is
indicated by their H$\alpha$ emission and has been confirmed by medium-resolution spectra,
the detection of lithium absorption and consistent RVs (also by the here presented 
observations).
The substellar nature is derived from the determination of bolometric
luminosities and effective temperatures (converted by means of temperature scales from
spectral types) and comparison with theoretical evolutionary tracks in the
Hertzsprung-Russell diagram (HRD).

Another substellar test applicable to young brown dwarfs was developed from a
test for old brown dwarfs based on lithium absorption (Rebolo et al. 1992)
to the age-independent statement, that \emph{any object with spectral type M7 or later that
shows lithium is substellar} (Basri 2000).
Since all Cha\,H$\alpha$ objects show lithium absorption in their spectra
(Comer\'on et al. 2000, Joergens \& Guenther 2001),
the four objects Cha\,H$\alpha$\,1, 7, 10 and 11 (M7.5--M8)
are thus bona fide brown dwarfs, 
with masses in the range of about 0.03\,M$_{\odot}$ to 0.05\,M$_{\odot}$,
Cha\,H$\alpha$\,2, 3, 6, 8, 9, 12 (M6.5--M7) can be classified as brown dwarf candidates
and Cha\,H$\alpha$\,4 and 5 (M6) are most likely very low-mass stars with 0.1\,M$_{\odot}$.
An error of one subclass in the determination of the
spectral type was taken into account.

We observed most of these ten brown dwarfs and brown dwarf candidates and seven (very) 
low-mass stars by means of high-resolution spectroscopy with UVES at the 8.2\,m telescope 
at the VLT and by means of a photometric monitoring campaign at a 1.5\,m telescope.  


\subsection{High-resolution UVES spectroscopy}
\label{sect:uves}

High-resolution spectra have been taken for the brown dwarfs and low-mass stars
Cha\,H$\alpha$\,1--8 and Cha\,H$\alpha$\,12, B34, CHXR\,74 and Sz\,23
between the years 2000 and 2004 with the cross-dispersed 
UV-Visual Echelle Spectrograph (UVES, Dekker\,et\,al.\,2000)
attached to the 8.2\,m Kueyen telescope of the Very Large Telescope (VLT) 
operated by the European Southern Observatory at Paranal, Chile.
The wavelength regime from 6600\,{\AA} to 10400\,{\AA} was covered
with a spectral resolution of $\rm \lambda / \Delta \lambda=40\,000$. 
For each object at least two spectra separated by a few weeks have been obtained
in order to monitor time dependence of the RVs.
For several objects, more than two and up to twelve spectra have been taken. 

After standard reduction, we have measured RVs by means of cross-correlating plenty of stellar
lines of the object spectra with a template spectrum.
In order to achieve a high wavelength and therefore RV precision, 
telluric O$_2$ lines have been used as wavelength reference.
A RV precision between 40\,m\,s$^{-1}$ and 
670\,m\,s$^{-1}$, depending on the S/N of the individual spectra, was achieved
for the relative RVs.
The errors are based on the standard deviation of two consecutive single spectra.
An additional error of about 300\,m\,s$^{-1}$ has to be taken into account for the 
absolute RVs due to uncertainties in the zero point of the template.
These RVs are one of the most precise ones for young brown dwarfs and very low-mass stars
available up to now.

Based on the time-resolved RVs, a RV survey for planetary and brown dwarf
companions to the targets was carried out (Sect.\,\ref{sect:rvsurvey}).
The mean RVs were explored in a kinematic study of this group of brown dwarfs
in Sect.\,\ref{sect:kinematic}.
Additionally, projected rotational velocities $v \sin i$
were derived from line-broadening of spectral features (Sect.\,\ref{sect:rotation}), 
lithium equivalent width has been measured to confirm  
youth and membership to the Cha\,I star forming region
and the CaII IR-triplet has been studied as an indicator for chromospheric activity.
Results have been or will be published in Joergens \& Guenther (2001), Joergens (2003),
Joergens et al. (2005a,b).

\subsection{Photometric monitoring}
\label{sect:phot}

In order to study the time-dependent photometric behaviour of the brown dwarfs
and very low-mass stars,
we monitored a 13$^{\prime} \times 13^{\prime}$ region in the Cha\,I cloud
photometrically in the Bessel R and the Gunn i filter in six consecutive half
nights with the CCD camera DFOSC at the Danish 1.5\,m telescope at La Silla, 
Chile in May and June 2000.
Differential magnitudes were determined relative to a set of reference stars in the same field 
by means of aperture photometry.
The field contained all of the targets introduced in the previous sections, however,
CHXR\,74 and Sz\,23 were too bright and have been saturated in the images,
while Cha\,H$\alpha$\,1, 7, 9, 10 and 11 were too faint for the size of the telescope.

Based on the lightcurves for Cha\,H$\alpha$\,2, 3, 4, 5, 6, 8, 12,  B\,34, CHXR\,78C 
and CHXR\,73, we searched for periodic variabilities and were able to 
determine absolute rotation periods for several brown dwarfs and very low-mass stars
based on brightness modulations due to surface activity (Sect.\,\ref{sect:rotation},
Joergens et al. 2003b).

\section{Kinematics of brown dwarfs in Cha\,I \\ and the ejection scenario}
\label{sect:kinematic}

It was proposed that brown dwarfs might form in a star-like manner
but have been prevented from accreting to stellar masses by the early ejection 
out of their birth place (Sect.\,\ref{sect:formation}).
The ejection process might have left an observable imprint in the kinematics of 
ejected members of a cluster in comparison to that of non-ejected members.
Therefore, we are studying the kinematics of our target brown dwarfs in Cha\,I
by means of precise mean RVs measured from UVES spectra.
We find that the seven brown dwarfs and brown dwarf candidates and two
0.1\,M$_{\odot}$ stars in Cha\,I (spectral types M6--M8)
differ very less from each other in terms of their RV.
The sample has a mean RV of 15.7\,km\,s$^{-1}$ and a RV dispersion measured in terms of
standard deviation of only 0.9\,km\,s$^{-1}$. The total covered RV range is 2.6\,km\,s$^{-1}$. 
We note, that in previous publications (Joergens \& Guenther 2001, Joergens 2003),
the dispersion was measured in terms of full width at half maximum (fwhm)
of a Gaussian distribution 
(which  is related to the standard deviation 
$\sigma$ of the Gaussian by fwhm\,=\,$ \sigma \sqrt{8 \ln 2}$) following the procedure 
of radio astronomers. 
Based on new RV data and an improved data analysis, the here presented kinematic study 
(Joergens et al. 2005b) is a revised version of the previous ones.
The new fwhm for the brown dwarfs is 2.1\,km\,s$^{-1}$, which is 
consistent with the previous value of 2.0\,km\,s$^{-1}$.

\begin{figure}[t]
\begin{center}
\includegraphics[width=.55\textwidth,angle=270,clip]{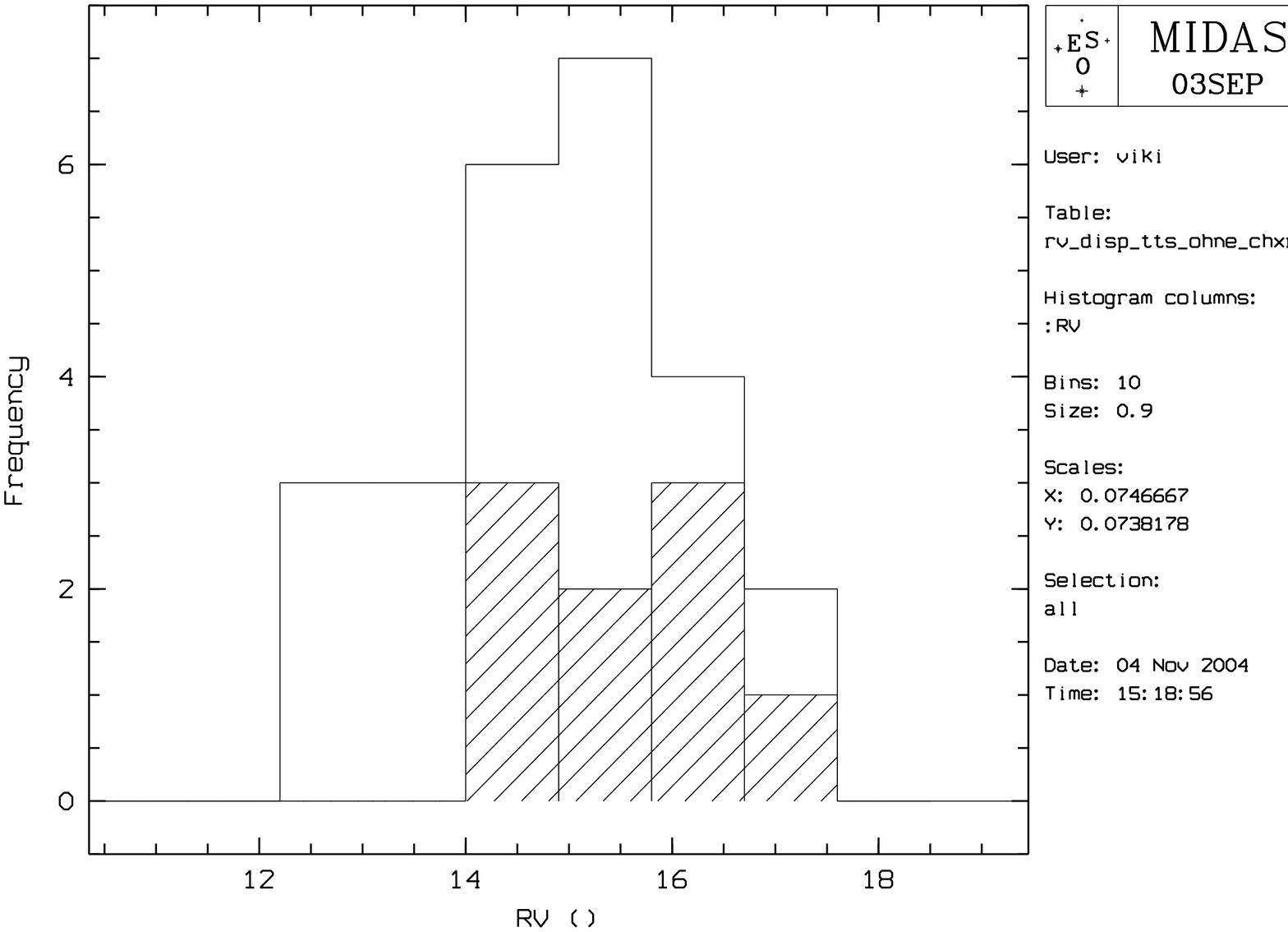}
\end{center}
\caption
[RV distribution of brown dwarfs in Cha\,I]
{\label{fig:kinematics}
\small{{\bf Distribution of mean RVs} in km\,s$^{-1}$ for nine 
brown dwarfs and very low-mass stars with masses $\leq$ 0.1\,M$_{\odot}$ 
(hashed) and for 25 T~Tauri stars in Cha\,I.
}}
\label{hist}
\end{figure}
%

In order to compare the results for the brown dwarfs in Cha\,I with the 
RV distribution of higher-mass stellar objects in this cluster,
we compiled all T~Tauri stars confined to the same region 
for which RVs have been measured with a precision of 2\,km\,s$^{-1}$ or better
from the literature 
(Walter 1992, Dubath et al. 1996, Covino et al. 1997, Neuh\"auser \& Comer\'on 1999), 
from Guenther et al. (in prep., see Joergens \& Guenther 2001)
and from our own measurements based on UVES spectra for M-type T~Tauri stars.
The compiled 25 T~Tauri stars (spectral types M5--G2)
have a mean RV of 14.7\,km\,s$^{-1}$ and a RV dispersion in terms of standard deviation of
1.3\,km\,s$^{-1}$. The dispersion given in terms of fwhm is 3.1\,km\,s$^{-1}$.
Compared to the previous studies (Joergens \& Guenther 2001, Joergens 2003),
the T~Tauri sample and RVs have been also revised by taking into account new UVES-based 
RVs obtained by us in 2002 and 2004 as well as overlooked RV measurements by Walter (1992). 
Furthermore, based on a recent census of Cha\,I (Luhman 2004)
an up-to-date check of membership status was possible.
The new standard deviation of 1.3\,km\,s$^{-1}$ differs only slightly from the previous one
of 1.5\,km\,s$^{-1}$. 

The resulting RV distributions of the T~Tauri stars and the brown dwarfs
are displayed in Fig.\,\ref{fig:kinematics} in form of a histogram.
The mean RV of the M6--M8 type brown dwarfs and very low-mass stars
(15.7\,km\,s$^{-1}$) is consistent within the errors with that of the T~Tauri stars
in the same field (14.7\,km\,s$^{-1}$) and 
with that of the molecular gas of the surrounding (15.3\,km\,s$^{-1}$, Mizuno et al. 1999).
This confirms the membership of the substellar population to the Cha\,I star forming cloud.

Furthermore, we find that the RV dispersion of the brown dwarfs (0.9\,km\,s$^{-1}$) 
is slightly smaller than that of the
T~Tauri stars in the same field (1.3\,km\,s$^{-1}$) and slightly larger than
that of the surrounding molecular gas 
(the fwhm of 1.2\,km\,s$^{-1}$ given by Mizuno et al. (1999) translates to a 
standard deviation of 0.5\,km\,s$^{-1}$ using the above given formula).
For the comparison of the kinematic properties of brown dwarfs and stellar objects we have 
set the dividing line between the two populations arbitrarily at 0.1\,M$_{\odot}$ knowing 
that this is not exactly the substellar border. However, the process forming brown dwarfs 
is expected to work continuously into the regime of very low-mass stars
and it is a priori unknown if and for which mass range a different formation mechanism is 
operating. In a more detailed study (Joergens et al 2005b), we calculate
the dispersion also for different choices of this dividing line but see no significant
differences in the results. 

RVs are tracing only space motions in the direction of the line-of-sight and the 
studied brown dwarfs could have a larger three-dimensional velocity dispersion.
However, the studied brown dwarfs in Cha\,I have an age of 1--5\,Myr 
and occupy a field of less than
12$^{\prime}$\,$\times$\,12$^{\prime}$ at a distance of 160\,pc.
Brown dwarfs born within this field and ejected during the early accretion phase in 
directions with a significant fraction perpendicular to the line-of-sight, would have flown 
out of the field a long time ago for velocities of 0.5\,km\,s$^{-1}$ or larger
(Joergens et al. 2003a).
Therefore, the measurement of RVs is a suitable method to
test if objects born and still residing in this field have significantly high velocities due to
dynamical interactions during their formation process.

Numerical simulations of brown dwarf formation by the ejection scenario
differ in their predictions of the resulting velocities over a wide range.
In general, the ejection velocity scales with the inverse of the square root of the distance 
of the closest approach in the encounter that led to the ejection (Armitage \& Clarke 1997).
The resulting velocity distribution depends also on the gravitational potential of the 
cluster since it defines if an ejected object with a certain velocity can escape or is 
ejected only in a wide eccentric orbit (e.g. Kroupa \& Bouvier 2003). 
Furthermore, the kinematic signature of the ejection can be washed out 
if only part of the brown dwarfs in a cluster are formed according to the ejection scenario
while a significant fraction is 
formed by other mechanisms (disk instabilities alone 
or direct collapse above the fragmentation limit), or when, on the other hand,
all (sub)stellar objects in a cluster undergo significant dynamical interactions.

Hydrodynamic calculations by Bate et al. (2003) and Bate \& Bonnell (2005)
predict velocity dispersions of 
2.1\,km\,s$^{-1}$ and 4.3\,km\,s$^{-1}$, resp., and no kinematic difference between  
brown dwarfs and T~Tauri stars.
These calculations are performed for much denser star forming regions than Cha\,I and an extrapolation 
of their results to the low-density Cha\,I cloud might lead to a consistency with our observed RV dispersion 
of 0.9\,km\,s$^{-1}$ for brown dwarfs and only slightly larger for T~Tauri stars.
However, comparison with Delgado-Donate et al. (2004) indicates that the dependence
of the velocity dispersion on the stellar density is not yet well established. 
N-body simulations by Sterzik \& Durisen (2003) predict that 25\% of
brown dwarf singles have a velocity smaller than 1\,km\,s$^{-1}$, that is much lower
than our observations of 67\% of brown dwarfs in Cha\,I having RVs smaller than 1\,km\,s$^{-1}$.
Also the high velocity tail found by the authors of 
40\% single brown dwarfs having higher velocities than 1.4\,km\,s$^{-1}$ and 10\% $>$5\,km\,s$^{-1}$  
is not seen in our data, were none has a RV deviating by the mean by more than
1.4\,km\,s$^{-1}$.
Recent N-body calculations by Umbreit et al. (2005) showed that
the ejection velocities depend strongly on accretion
and by assuming different accretion models and rates, the authors predict an even more 
pronounced high-velocity tail with  
60\% to 80\% single brown dwarfs having velocities larger than 1\,km\,s$^{-1}$.
This is also much larger than found by our observations, where only about 30\% have
velocities $>$ 1\,km\,s$^{-1}$. 
One might argue that the non-detection of a high velocity tail in our data can be attributed to 
the relatively small size of our sample, on the other hand, the current N-body simulations do not 
take into account the gravitational potential of the cluster, which might cause a suppression  
of the highest velocities.

To conclude, the observed values of our kinematic study provide the first
observational constraints for the velocity distribution of a group of very young 
brown dwarfs and show that they have a RV dispersion of 0.9\,km\,s$^{-1}$, no high-velocity tail
and their RVs are not more dispersed than that of T~Tauri stars in the same field.
These observed velocities are smaller than any of the theoretical predictions for brown dwarfs formed
by the ejection scenario, which might be attributed to 
the lower densities in Cha\,I compared to some model assumptions, to shortcomings in the models, like 
neglection of feedback processes (Bate et al. 2003, Bate \& Bonnell 2005, Delgado-Donate et al. 2004) 
or of the cluster potential (Sterzik \& Durisen 2003, Umbreit et al. 2005), 
or to the fact that our sample is statistically relatively small.
The current conclusion is that either the brown dwarfs in 
Cha\,I have been formed by ejection but with smaller velocities as theoretically predicted 
or they have not been formed in that way.

\section{RV Survey for planets and \\brown dwarf companions with UVES}
\label{sect:rvsurvey}

Based on the precise RV measurements in time resolved UVES spectra (Sect.\,\ref{sect:uves}),
we are carrying out a RV survey for (planetary and brown dwarf) companions to the young brown
dwarfs and very low-mass stars in the Cha\,I cloud.

The detection of planets around brown dwarfs as well as the detection of
young spectroscopic brown dwarf binaries
would be an important clue towards the formation of brown dwarfs.
So far, no planet is known orbiting a brown dwarf (the recent publication
of a direct image of a candidate for a  5\,M$_\mathrm{Jup}$ planet around a 
25\,M$_\mathrm{Jup}$ brown dwarf by Chauvin et al. (2004) is very exciting but 
still very tentative since it might very well be a background object).
There have been detected several brown dwarf binaries, among them
there are three spectroscopic, and hence close systems
(Basri \& Mart\'{\i}n 1999, Guenther \& Wuchterl 2003).
However, all known brown dwarf binaries are fairly old and
it is not yet established if the typical outcome of
the brown dwarf formation process is a binary or multiple brown dwarf system
or a single brown dwarf.

Furthermore, the search for planets around very young as well as around very low-mass stars
and brown dwarfs is interesting since
the detection of \emph{young} planets as well as a census of planets around
stars of all spectral types, and maybe even around brown dwarfs, is an important step
towards the understanding of planet formation. It would provide empirical
constraints for planet formation time scales. Furthermore, it would show if planets can 
exist around objects which are of considerably lower mass and surface
temperature than our sun.

\subsection{RV constant objects}

\begin{figure}[t]
\centering
\includegraphics[width=0.9\linewidth]{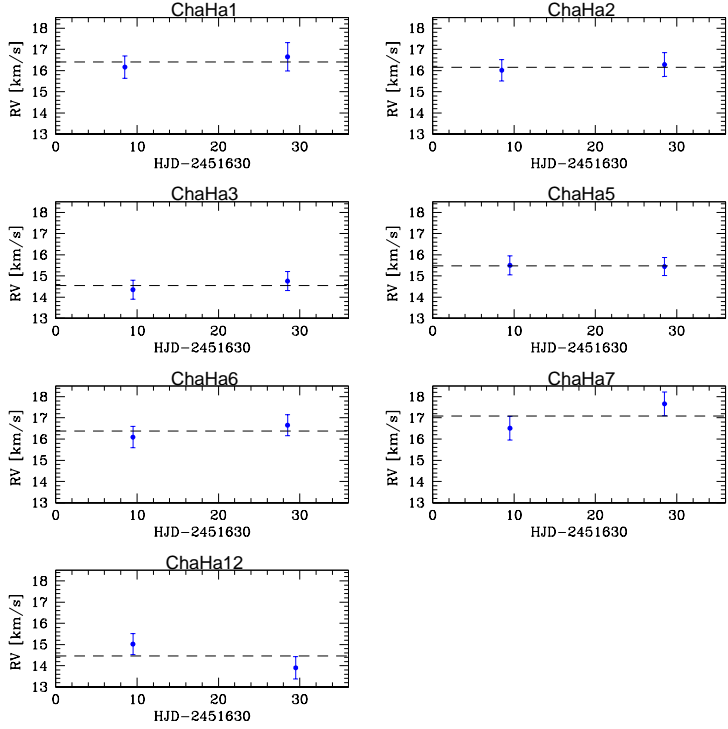}
\caption{
\label{fig:rv1a}
RV constant objects:
RV vs. time in Julian days for brown dwarfs and very low-mass stars 
(M6--M8)
in Cha\,I based on high-resolution UVES/VLT spectra.
Error bars indicate 1 $\sigma$ errors.
}
\end{figure}

\begin{figure}[t]
\centering
\includegraphics[width=0.5\linewidth]{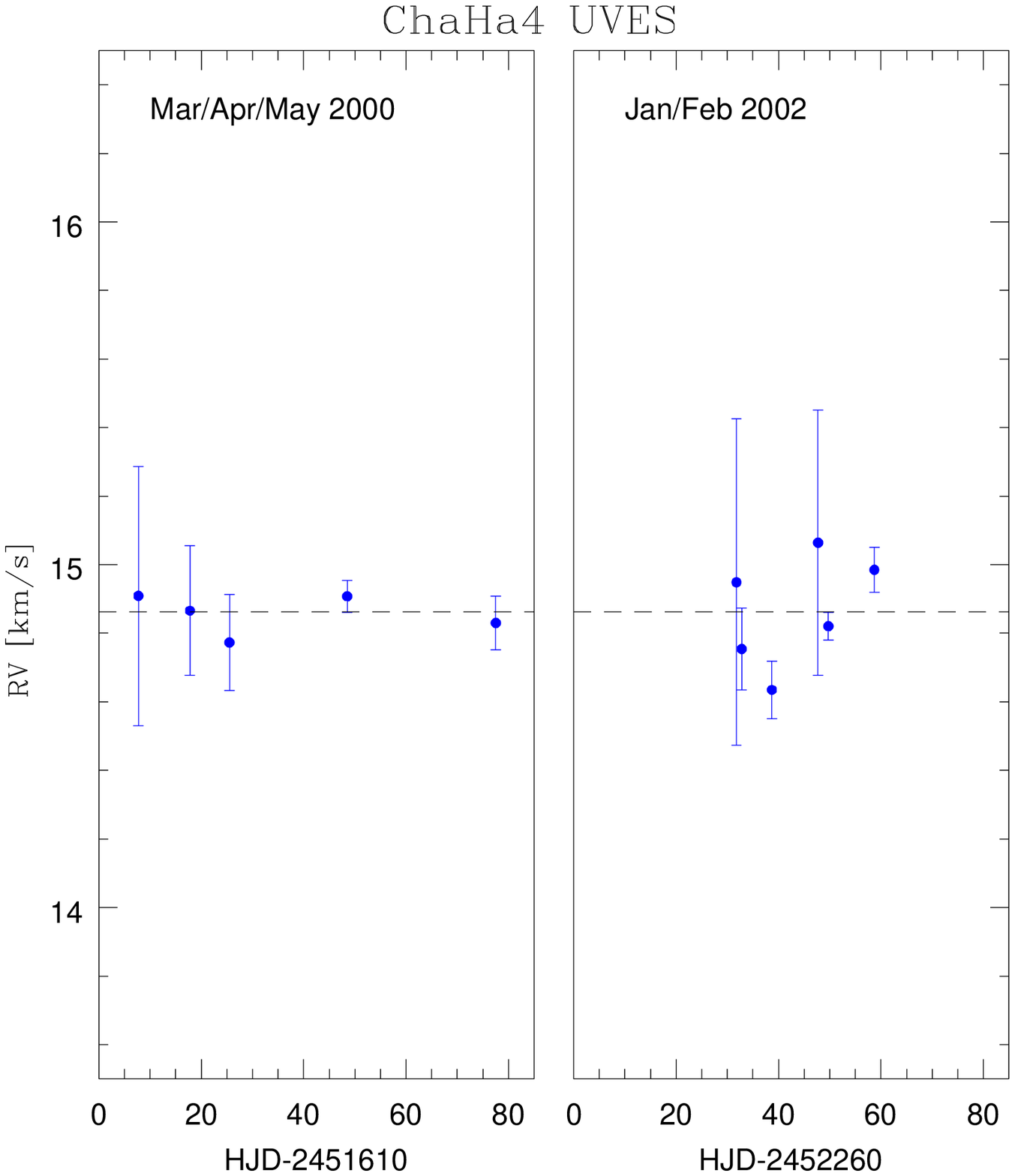}
\includegraphics[width=0.5\linewidth]{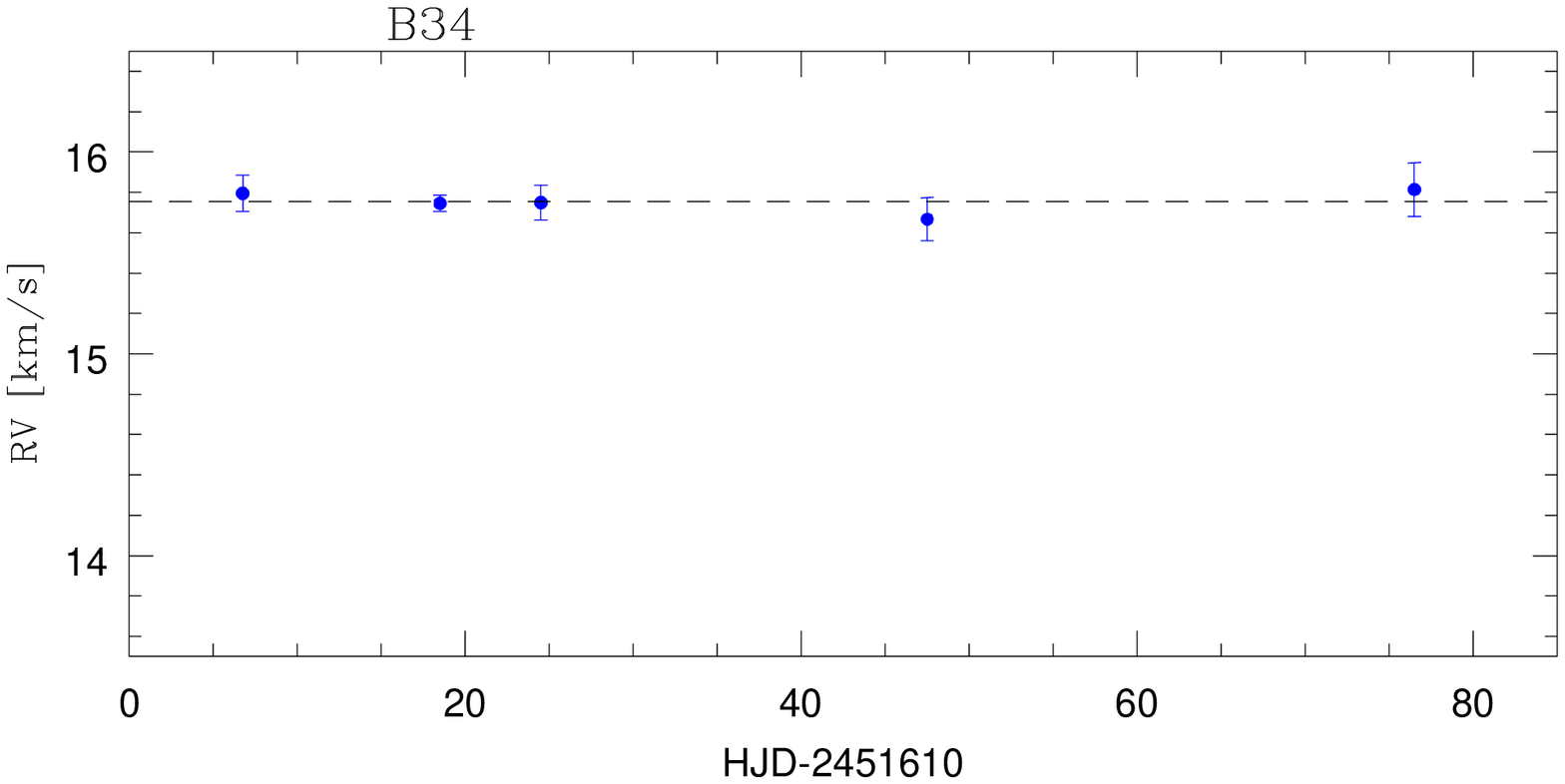}
\caption{
\label{fig:rv1b}
RV constant objects (continued): RV vs. time in Julian days for very low-mass stars 
($\sim$0.1\,M$_{\odot}$) in Cha\,I based on high-resolution UVES/VLT spectra.
Error bars indicate 1 $\sigma$ errors.
}
\end{figure}

The RVs for the brown dwarfs and very low-mass stars
Cha\,H$\alpha$\,1, 2, 3, 4, 5, 6, 7, 12 and B\,34 are constant within the 
measurements uncertainties of 2\,$\sigma$ for Cha\,H$\alpha$\,4 and of 1\,$\sigma$ for 
all others, as displayed in Figs.\,\ref{fig:rv1a}, \ref{fig:rv1b}.  
From the non-detections of variability, we have estimated upper limits for 
the projected masses $M_2 \sin i$ of hypothetical companions for each 
object\footnote{Spectroscopic detections of companions allow in general no absolute mass 
determination but only the derivation of a lower limit of the companion mass
$M_2 \sin i$ due to the unknown inclination $i$.}. 
They range between 0.1\,M$_\mathrm{Jup}$ and 1.5\,M$_\mathrm{Jup}$ assuming a 
circular orbit, a separation of 0.1\,AU between companion and primary object and
adopting primary masses from Comer\'on et al. (1999, 2000).
The used orbital separation of 0.1\,AU corresponds to
orbital periods ranging between 30 and 70 days for the masses of these
brown dwarfs and very low-mass stars. 

That means, that these nine brown dwarfs and very low-mass stars with spectral types 
M5--M8 and masses $\leq$ 0.12\,M$_{\odot}$ show no RV variability down to Jupitermass planets.
There is, of course, the possibility that present companions have not been detected
due to non-observations at the critical orbital phases.
Furthermore, long-period companions may have been missed since for all of them but 
Cha\,H$\alpha$\,4, the time base of the observations does not exceed two months.

\subsection{RV variable objects}

\begin{figure}[t]
\centering
\includegraphics[width=0.48\linewidth]{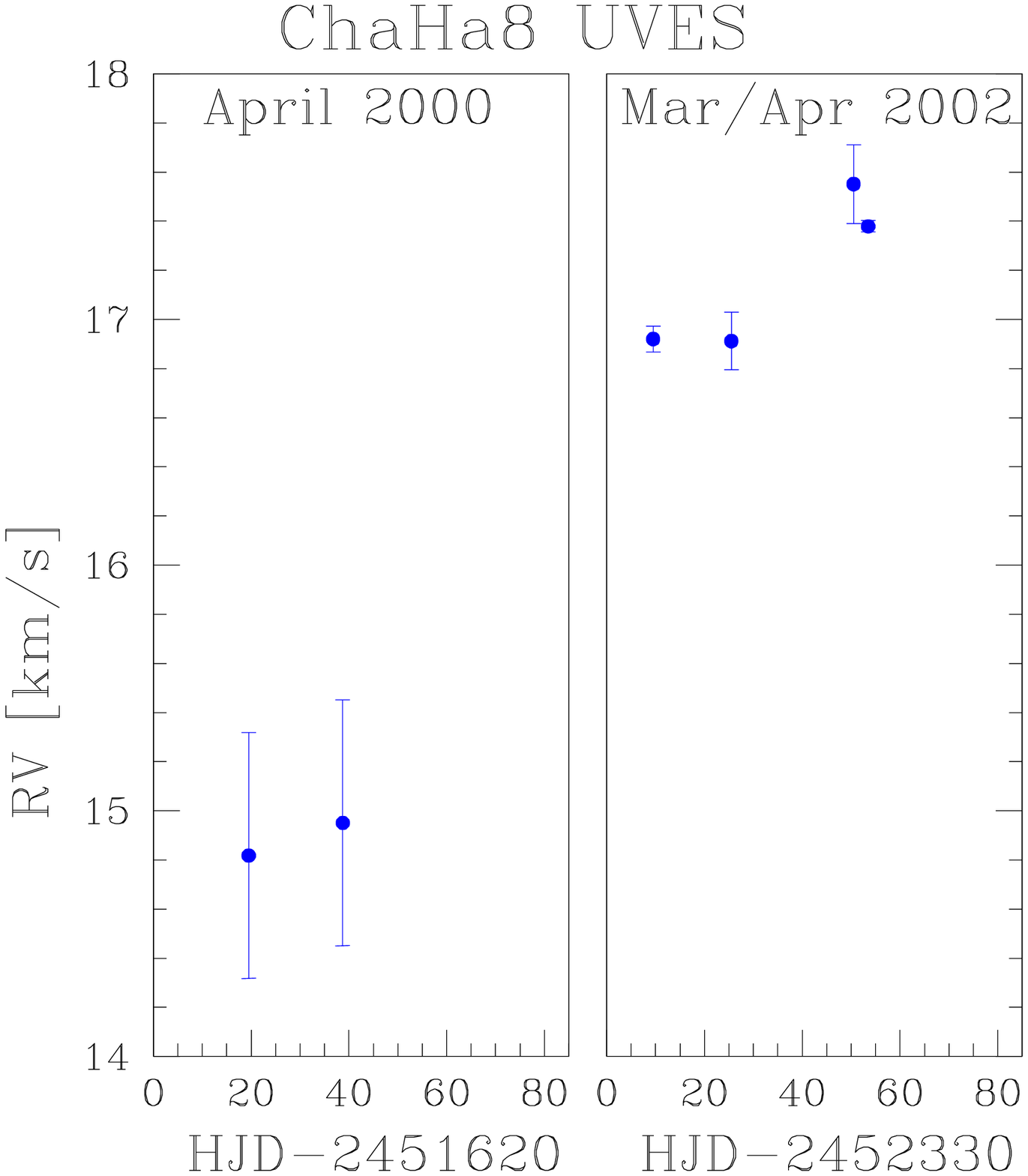}
\includegraphics[width=0.48\linewidth]{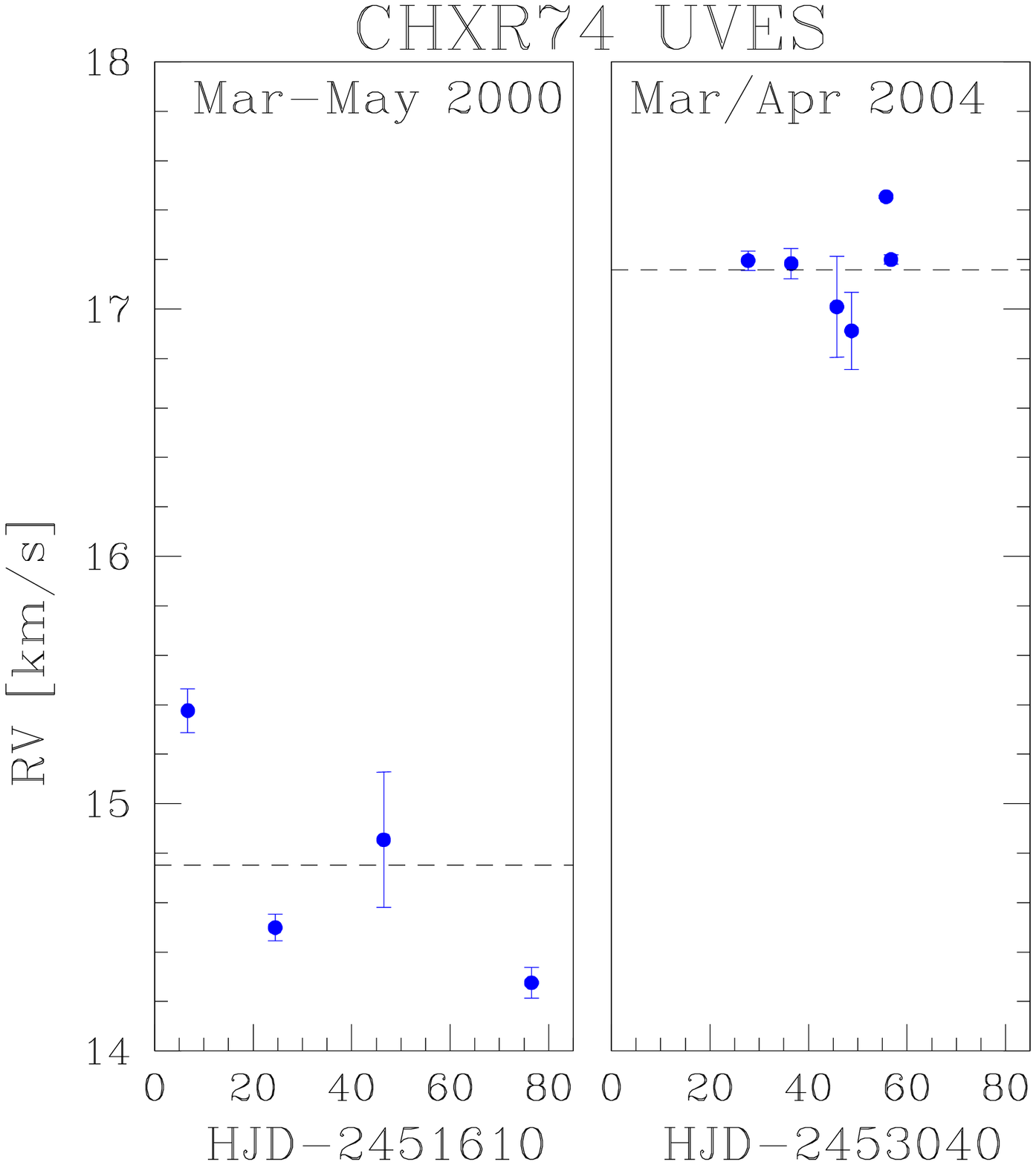}
\includegraphics[width=0.48\linewidth]{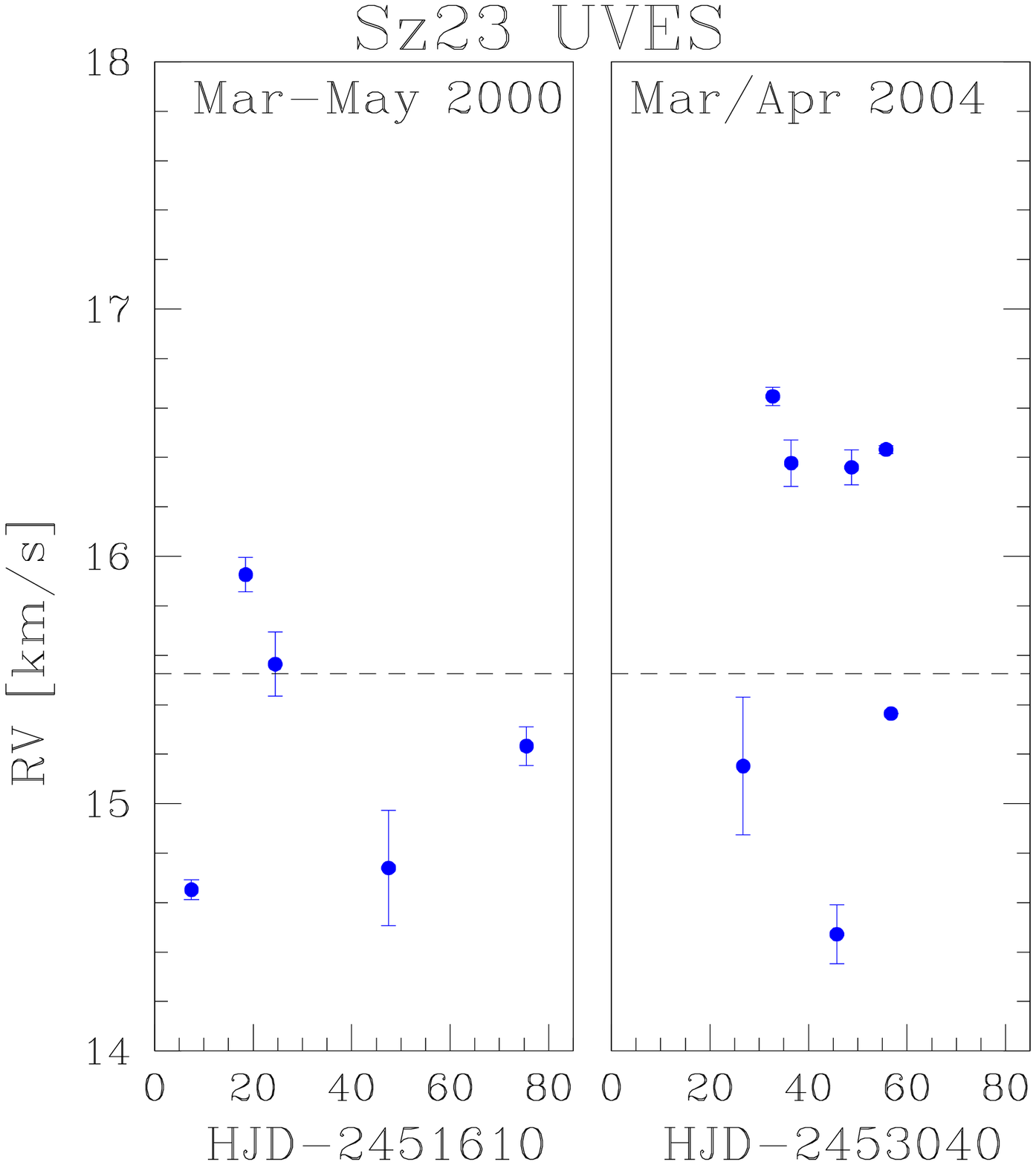}
\caption{
\label{fig:rv2}
RV variable objects: RV vs. time in Julian days for the brown dwarf candidate
Cha\,Ha\,8 (M6.5) and the low-mass T~Tauri stars
CHXR\,74 and Sz\,23 ($\sim$0.2--0.3\,M$_{\odot}$) based on high-resolution UVES/VLT spectra.
Error bars indicate 1\,$\sigma$ errors.
}
\end{figure}

For three objects, we have found significant RV variations,
namely for the brown dwarf Cha\,H$\alpha$\,8 and 
the low-mass stars CHXR\,74 ($\sim$0.17\,M$_{\odot}$) and
Sz\,23 ($\sim$0.3\,M$_{\odot}$) as shown in Fig.\,\ref{fig:rv2}.
The variability characteristic differs among the three objects.
Sz\,23 shows variability on time scales of days with no difference 
in the mean values of RVs recorded in 2000 and in 2004. 
On the other hand, Cha\,H$\alpha$\,8 and CHXR\,74
show only very small amplitude variations or no variations at all 
on time scales of days to weeks,
whereas the mean RV measured in 2000 differs significantly 
from the one measured years later, namely in 2002 for Cha\,H$\alpha$\,8 and 
in 2004 for CHXR\,74, respectively, hinting at variability periods of
the order of months or longer.  

One  possibility for the nature of these RV variations is that they are the Dopplershift 
caused by the gravitational force of orbiting companions.
The poor sampling does not allow us to determine periods of the variations
but based on the data we suggest that the period for Cha\,H$\alpha$\,8 is 150 days or longer.
A 150\,d period would corresponds to a 6\,M$_\mathrm{Jup}$ 
planet orbiting at a separation of 0.2\,AU around Cha8. 
For longer periods the orbital separations as well as the 
mass of the companion would be larger.
For CHXR74, a period of about 200 days would be able to explain the 
RV data of 2000 and 2004 corresponds to a 15\,M$_\mathrm{Jup}$ 
brown dwarf orbiting CHXR74 at a separation of 0.4\,AU.

The other possibility is that they are caused by surface activity since prominent surface 
spots can cause a shifting of the photo center at the rotation period 
(see also Sect.\,\ref{sect:absrot}).
The upper limits for the rotational periods of Cha\,H$\alpha$\,8, CHXR\,74 and Sz\,23 
are 1.9\,d, 4.9\,d and 2.1\,d, respectively, based on projected rotational
velocities $v \sin i$ (Sect.\,\ref{sect:rotation}, Joergens \& Guenther 2001).
Thus, the time-scale of the RV variability of Sz\,23 is of the order 
of the rotation period and could be a rotation-induced phenomenon.
In contrast to the other RV variable objects, 
Sz\,23 is also displaying significant emission in the CaII IR triplet lines, 
which is an indicator for chromospheric activity.
A further study of time variations of these lines is underway.

The RV variability of Cha\,H$\alpha$\,8 and CHXR\,74 on time scales of months to years
cannot be explained by being rotational modulation. 
If caused by orbiting companions, the detected RV variations of CHXR\,74 and 
Cha\,H$\alpha$\,8 correspond to giant planets of a few Jupiter masses with periods of 
several months.

\subsection{Discussion: Multiplicity, separations, RV noise}

\begin{figure}[t]
\centering
\includegraphics[width=0.5\linewidth]{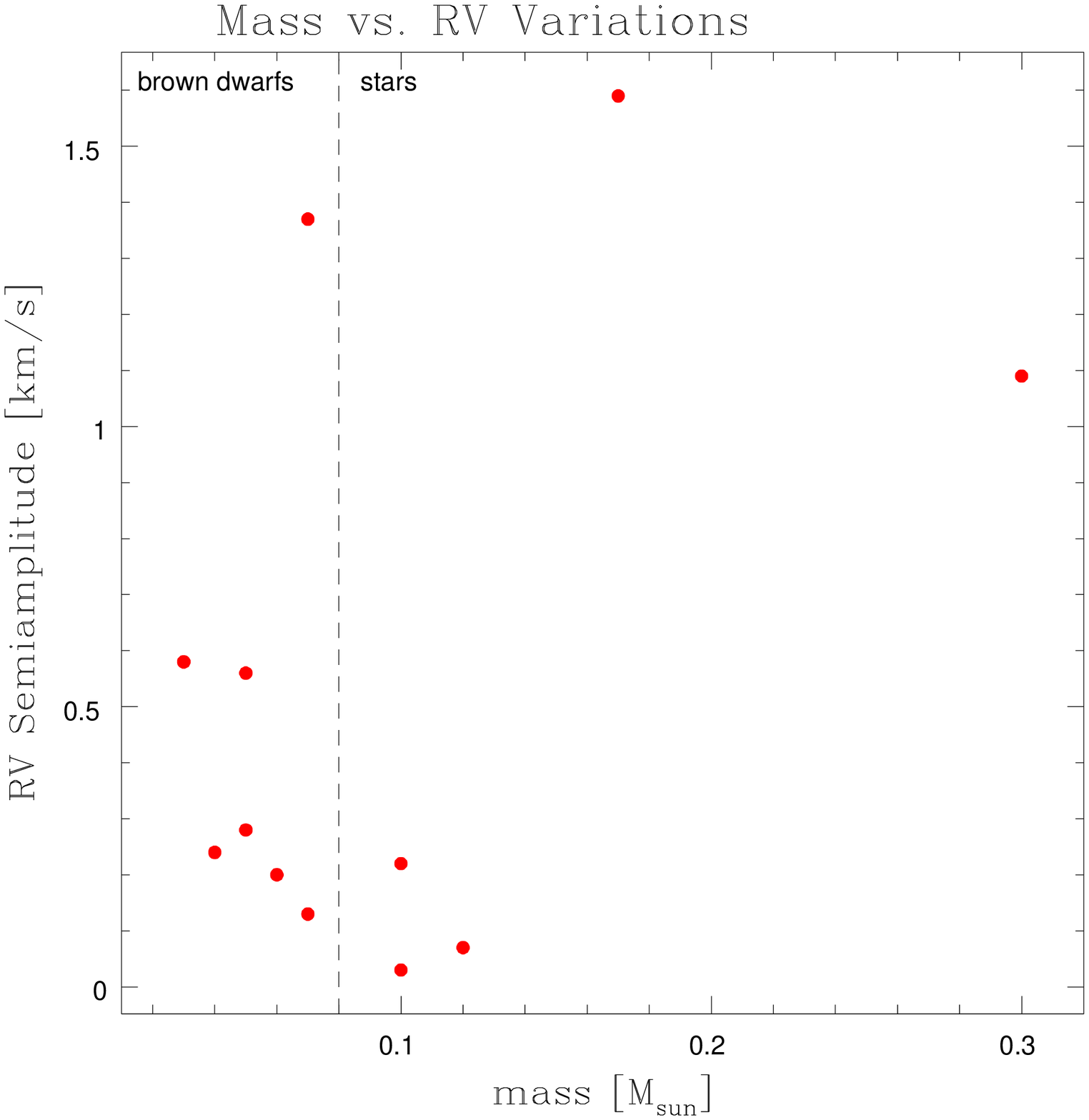}
\caption{
\label{fig:rvmass}
RV semiamplitude vs. object mass.
The upper three data points represent the RV variable objects with amplitudes above 
1\,km\,s$^{-1}$.
The remaining data points represent RV constant objects, which clearly follow 
a trend of decreasing RV amplitude with increasing mass. 
indicating that they are displaying no
significant RV noise due to surface spots down to the precision required to detect
Jupitermass planets, which would cause
and increasing RV amplitude with mass.
}
\end{figure}

Fig.\,\ref{fig:rvmass} displays for all targets of the RV survey the measured RV 
semiamplitude (for RV constant objects the upper limit for it) versus their mass,
as adopted from Comer\'on et al. (1999, 2000). 
The three RV variable objects have RV amplitudes above 1\,km\,s$^{-1}$ (top three data points 
in Fig.\,\ref{fig:rvmass}) and are clearly separated from the RV constant objects.
Interestingly, the RV constant objects follow a clear trend of decreasing RV amplitude 
with increasing mass. 
On one hand, this reflects simply the depending of the RV precision on the 
signal-to-noise of the spectra. On the other hand, it is an interesting finding by itself, 
since it shows that this group of brown dwarfs
and very low-mass stars with masses of 0.12\,M$_{\odot}$ and below display no
significant RV noise due to surface spots, which would cause
systematic RV errors with a RV amplitude increasing with mass.

\begin{figure}[t]
\centering
\includegraphics[width=0.7\linewidth]{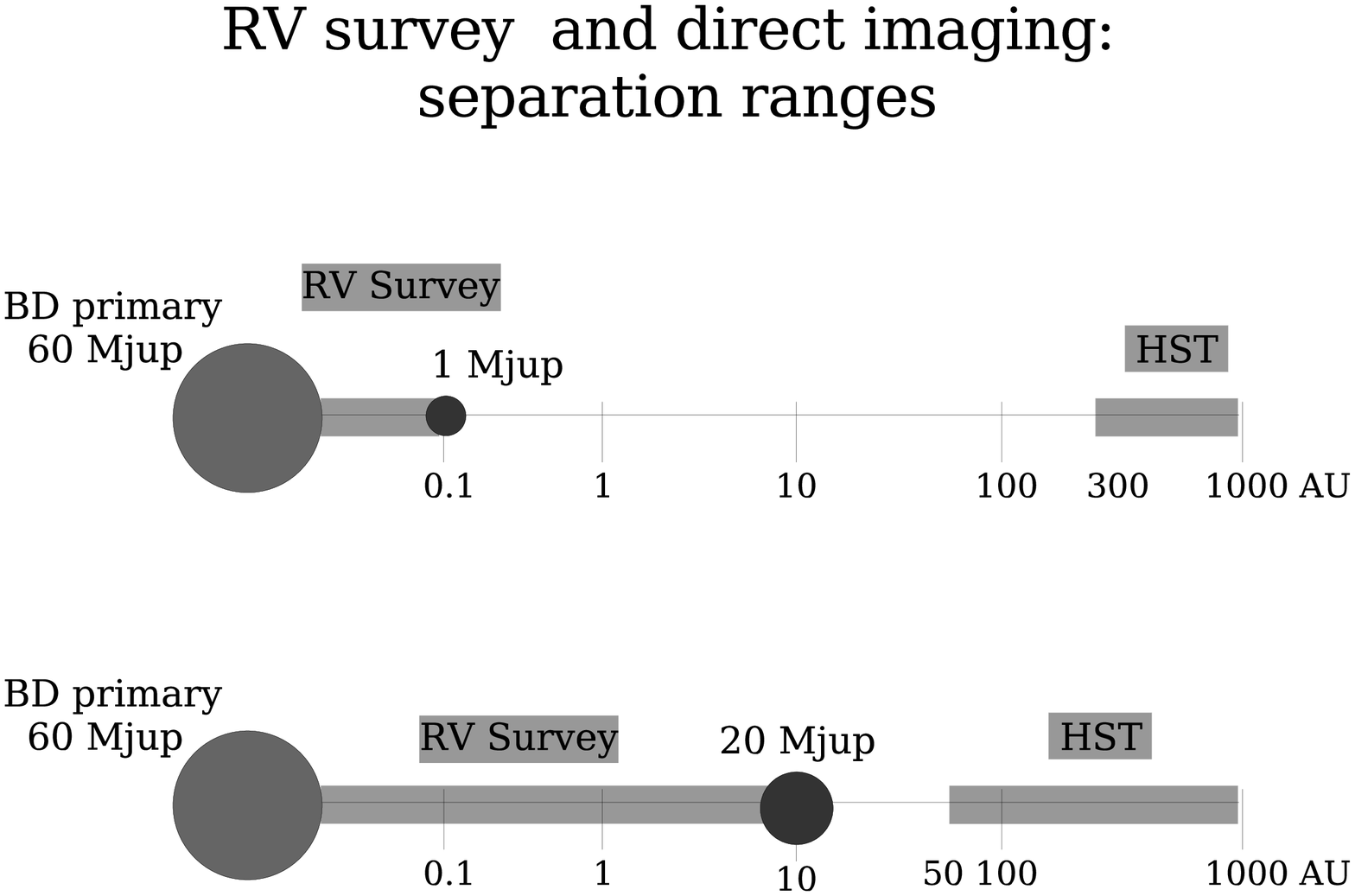}
\caption{
\label{fig:sep}
Separation ranges, which can be covered based 
on the achieved RV precision by our RV survey 
and by a direct imaging survey with the HST (Neuh\"auser et al. 2002).
An example is given for a one Jupitermass planet and a 20 Jupitermass brown dwarf orbiting 
a 60 Jupitermass brown dwarf.
Due to limited time base not the whole possible range for the RV survey has been covered yet.
}
\end{figure}

Among the subsample of ten brown dwarfs and very low-mass stars with masses $\leq$
0.12\,M$_{\odot}$
in this survey, only one (Cha\,H$\alpha$\,8) shows signs of RV variability, while the others
are RV quiet with respect to both companions and spots in our observations.
That hints at a very small multiplicity fraction of 10\% or less.
When considering also CHXR\,74, i.e. eleven objects (M$\leq$0.17\,M$_{\odot}$
and spectral types M4.5--M8), nine have constant RVs in the presented RV survey.
Interestingly, Cha\,H$\alpha$\,8 is RV constant and CHXR\,74 shows only small amplitude 
variations on time scales of days/weeks
but both reveal larger amplitude RV variability only on longer time scales of at least 
several months.
The fact that all other objects in this mass range do not show RV noise due to
activity suggests that the sources for RV variability due to activity are also weak for
Cha\,H$\alpha$\,8. Furthermore, the timescales of the variability are  
much too long for being caused by rotational modulation since the 
rotational period is of the order of 2 days.
The only other explanation would be a companion with a mass of several Jupitermasses
or more, i.e. a supergiant planet or a brown dwarf.
These observations give hints that companions to young brown dwarfs and very low-mass stars
might have periods of several months.

The RV survey probes the regions close to the central objects in respect of the occurrence 
of companions. Fig.\,\ref{fig:sep} shows the separation ranges, which can be covered based 
on the achieved RV precision. For example, a 20 Jupitermass brown dwarf in orbit around a 
60 Jupitermass brown dwarf would be detectable out to 10\,AU if the time base is long enough. 
For smaller companion masses the covered separation ranges are correspondingly smaller.
At the current stage, the limits in the covered separation range is set by the time base
rather than the RV precision. Therefore, further 3rd epoch RV measurements are planned.
The found small multiplicity fraction of the brown dwarfs and very low-mass stars 
in Cha\,I at small separations in this RV survey,
is also supported by the results of a direct imaging search for wide
(planetary or brown dwarf) companions to mostly the same targets, Cha\,H$\alpha$\,1--12, 
by Neuh\"auser et al. (2002, 2003), who find a multiplicity fraction of $\leq$10\%.
The separation ranges covered by this HST survey are also indicated in Fig.\,\ref{fig:sep}.

\section{Rotation}
\label{sect:rotation}

Measurements of rotation rates of young brown dwarfs are important 
to determine the evolution of angular momentum in the substellar 
regime in the first several million years of their lifetime,
during which rapid changes are expected due to the contraction on the 
Hayashi track, the onset of Deuterium burning and possible magnetic
interaction with a circumstellar disk.
Rotation speeds can be determined in terms of projected rotational velocities $v \sin i$
based on the line broadening of spectral features (Sect.\,\ref{sect:vsini})
or, if an object exhibits prominent surface features, which modulate the brightness
as the object rotates, the absolute rotation period can be 
determined by a light curve analysis (Sect.\,\ref{sect:absrot}). 
Both techniques have been applied to the brown dwarfs
and (very) low-mass stars in Cha\,I.

\subsection{Projected rotational velocities $v \sin i$}
\label{sect:vsini}

Projected rotational velocities $v \sin i$ have been measured 
based on the line broadening of spectral lines
in UVES spectra. 
We found that the $v\,\sin i$ values of the bona fide and candidate brown dwarfs in Cha\,I 
with spectral types M6--M8
range between 8\,km\,s$^{-1}$ and 26\,km\,s$^{-1}$ . The spectroscopic rotational 
velocities of the (very) low-mass stars Cha\,H$\alpha$\,4, Cha\,H$\alpha$\,5,  B\,34, 
CHXR\,74 and Sz\,23 are 14--18\,km\,s$^{-1}$ and, therefore, lie also within the range of 
that of the studied substellar objects.
These measurements provided the first determination of projected rotational velocities for 
very young brown dwarfs (Joergens \& Guenther 2001).
To compare them with $v\,\sin i$ values of older brown dwarfs, we consider
that the late-M type brown dwarfs in Cha\,I at and age of 1--5\,Myrs will 
further cool down and develop into L dwarfs at some point between and age of 
100\,Myr and 1\,Gyr (Burrows et al. 2001) and later into T dwarfs. 
$v\,\sin i$ values for old L dwarfs range between 10 and 60\,km\,s$^{-1}$ with the vast 
majority rotating faster than 20\,km\,s$^{-1}$
(Mohanty \& Basri 2003). 
Thus, the brown dwarfs in Cha\,I rotate on average slower than old L dwarfs in terms of 
rotational velocities $v \sin i$.
These results for brown dwarfs in Cha\,I are in agreement with $v\,\sin i$ values 
determined for five brown dwarf 
candidates in Taurus (7--14\,km\,s$^{-1}$, White \& Basri 2003) 
and for one brown dwarf in $\sigma$\,Ori, which has a $v\,\sin i$ of 9.4\,km\,s$^{-1}$ 
(Muzerolle et al. 2003) and an absolute rotational velocity of 14$\pm$4\,km\,s$^{-1}$
(Caballero et al. 2004).

Projected rotational velocities $v\,\sin i$ are lower limits of the rotational
velocity $v$ since the inclination $i$ of the rotation axis remains unknown.
Based on $v\,\sin i$ and the radius of the object an upper limit of the rotational period
P/$\sin i$ can be derived.
We estimated the radii of the brown dwarfs and very low-mass stars in Cha\,I
by means of the Stefan-Boltzmann law from bolometric 
luminosities and effective temperatures given by Comer\'on et al. (1999, 2000).
The approximate upper limits for their rotational periods 
range between one and three days for all studied brown dwarfs and (very) low-mass stars 
with the exception of the low-mass star CHXR\,74 (M4.5, $\sim$0.17\,M$_{\odot}$), for which
P/$\sin i$ is five days.

\subsection{Absolute rotational periods from lightcurve \\modulations}
\label{sect:absrot}

\begin{figure}[h]
\begin{center}
\includegraphics[width=.5\textwidth]{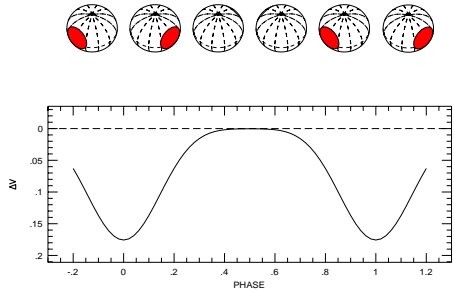}
\end{center}
\caption[Modulation of the brightness due to a spot]
{\label{spot}
\small{{\bf Rotational brightness modulation caused by surface spot.}
Courtesy of G. Torres.
A cool spot on the surface of a star\,/\,brown dwarf causes
a periodic dimming of the total brightness at the rotation period.
Plotted are the relative V magnitudes $\Delta$V over the orbital phase.
}}
\end{figure}

An object exhibiting prominent surface features distributed inhomogeneously over
its photosphere provides a way to measure its absolute rotation period
since the surface spot(s) cause a periodic
modulation of the brightness of the object as it rotates.
See Fig.\,\ref{spot} for an illustration.

\begin{figure}[t]
\begin{center}
\includegraphics[width=.49\textwidth]{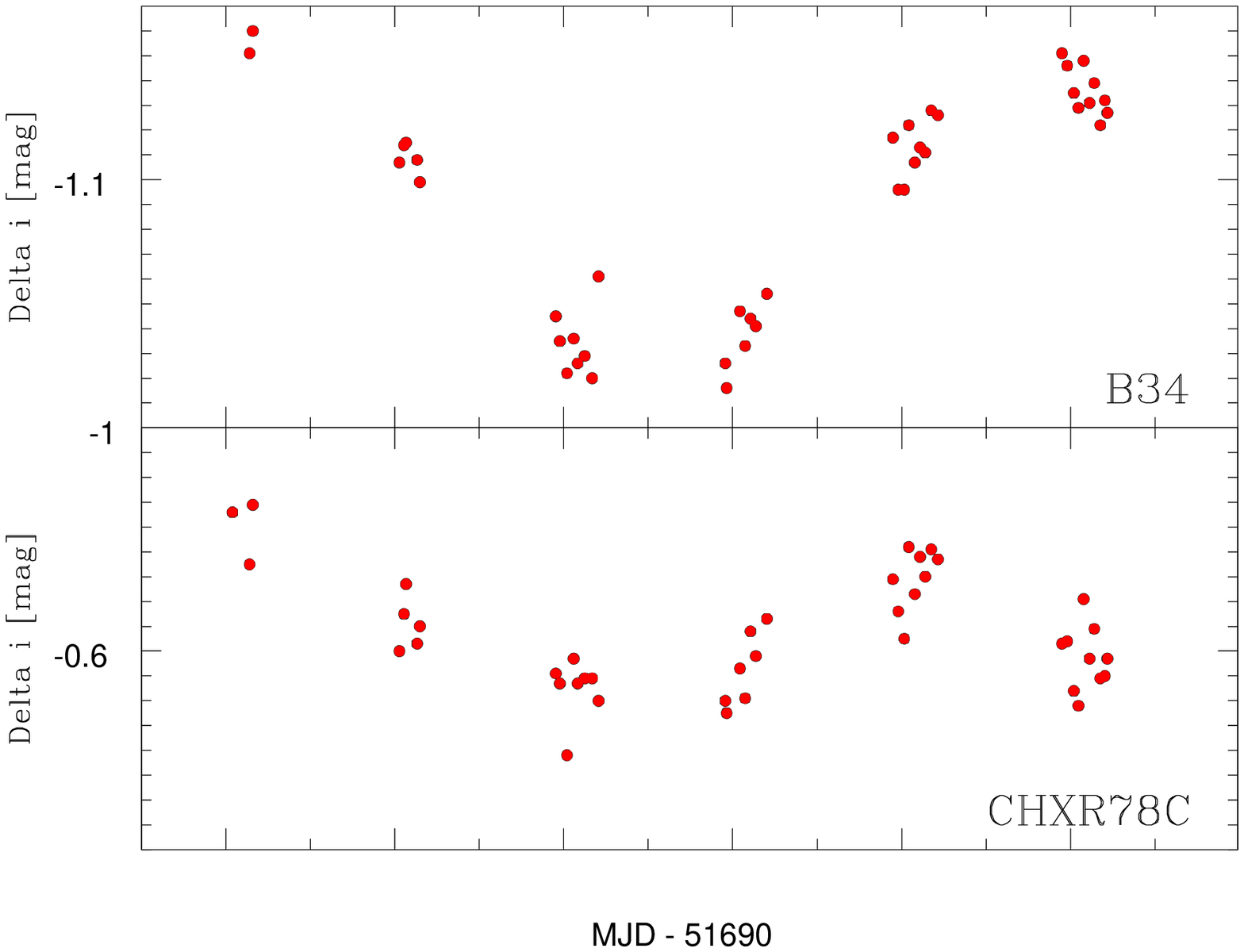}
\includegraphics[width=.49\textwidth]{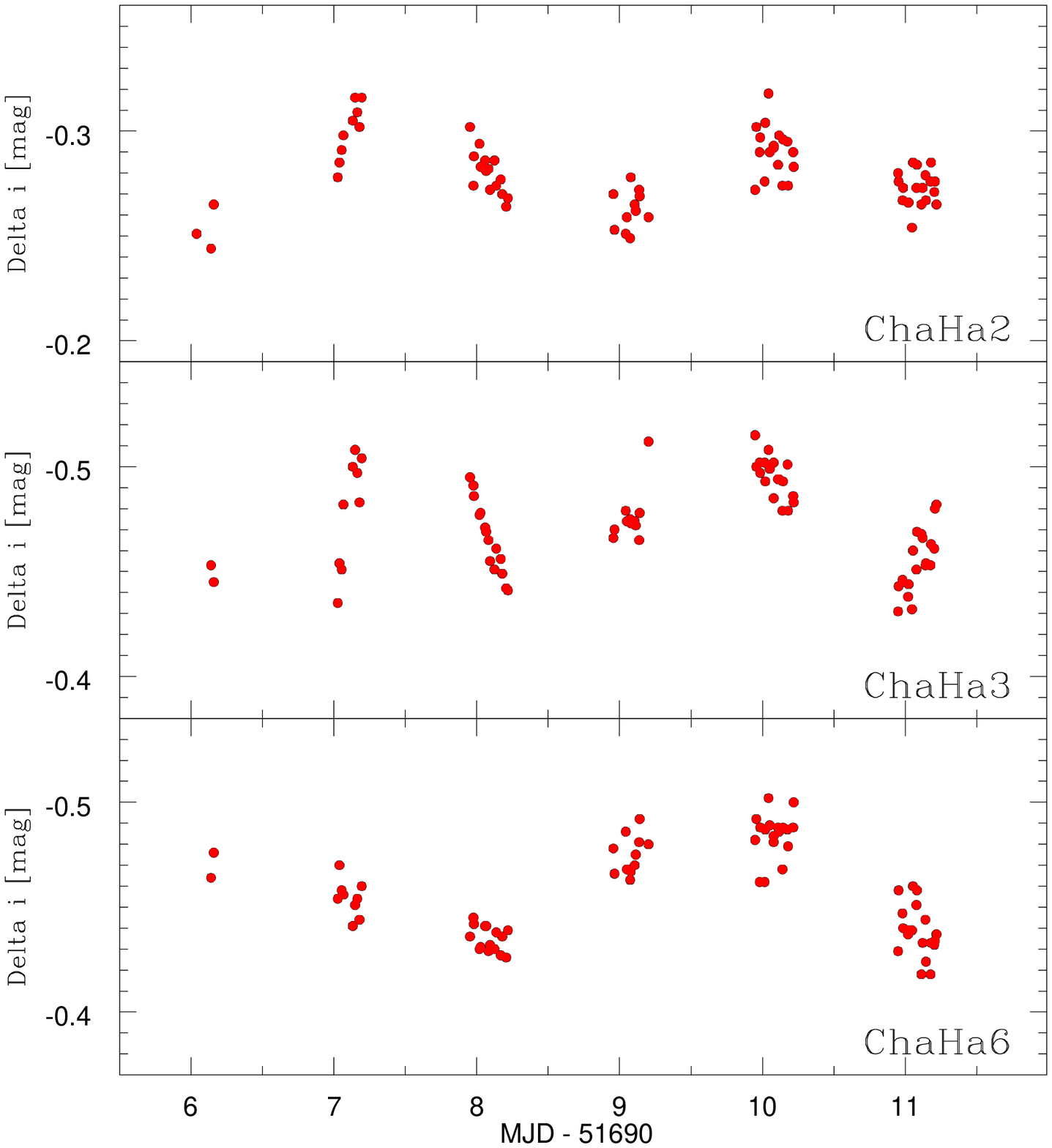}
\end{center}
\caption
[Original light curves for two very low-mass stars (left panel)
and three brown dwarf candidates (right panel)]
{\label{lcs}
\small{{\bf Light curves for two very low-mass stars (left panel) and three brown
dwarf candidates (right panel).}
From top to bottom and left to right are displayed i-band light curves for the
very low-mass stars B34 (M5, 0.12\,M$_{\odot}$) with a period of 4.5 days
and CHXR78C (M5.5, 0.09\,M$_{\odot}$) with a period of 3.9 days,
and for the three brown dwarf candidates Cha\,H$\alpha$\,2 (M6.5, 0.07\,M$_{\odot}$)
with a period of 2.8 days,
Cha\,H$\alpha$\,3 (M7, 0.06\,M$_{\odot}$) with a period of 2.2 days
and Cha\,H$\alpha$\,6 (M7, 0.05\,M$_{\odot}$) with a period of 3.4 days.
}}
\end{figure}

Based on photometric observations in the R and i band filter
(see Sect.\,\ref{sect:phot}), we searched for 
periodic variations in the light curves of the targets with 
the string-length method (Dworetsky 1983).
Periodic brightness variations were found for the three brown dwarf candidates
Cha\,H$\alpha$\,2, Cha\,H$\alpha$\,3, Cha\,H$\alpha$\,6
and two very low-mass stars B\,34 and CHXR\,78C. 
The original light curves are shown in Fig.\,\ref{lcs}.
In addition to i and R band data, we have analysed J-band monitoring data of the targets 
(Carpenter et al. 2002), which have been taken a few weeks earlier
and confirm the periods found in the optical data.
The determined periods are interpreted as rotation periods based on a consistency 
with the $v \sin i$ values from UVES spectra (previous section), 
recorded color variations in agreement with the expectations for spots on the surface
and the fact that with effective temperatures of more than 2800\,K, 
the objects are young and still hot and, therefore, their
atmospheric gas is very likely sufficiently ionized for the formation of spots.
Additionally, variability due to clouds, which could occur on time scales of
the formation and evolution of clouds,
can be excluded because their temperatures are too high
for significant dust condensation
(e.g. Tsuji et al. 1996a,b, Allard et al. 1997, Burrows \& Sharp 1999).

The found rotational periods of the three brown dwarf candidates
Cha\,H$\alpha$\,2, Cha\,H$\alpha$\,3 and Cha\,H$\alpha$\,6
are 3.2\,d, 2.2\,d and 3.4\,d, respectively (Joergens et al. 2003b).
The results show that brown dwarfs at an age of 1--5\,Myr display an inhomogeneous 
surface structure and rotate slower than old brown dwarfs (rotational periods below 
one day, e.g. Bailer-Jones \& Mundt 2001, Mart\'\i n et al. 2001). 
It is known that Cha\,H$\alpha$\,2 and 6 have optically thick disks 
(Comer\'on et al. 2000), therefore magnetic braking due to 
interactions with the disk may play a role for them.
This is suggested by the fact, that among the three brown dwarf candidates with determined 
periods, the one without a detected disk, Cha\,H$\alpha$\,3, has the shortest period.

\subsection{Rotation of young brown dwarfs}

The brown dwarfs in Cha\,I have rotation periods of 2--3 days
and projected rotation velocities $v \sin i$ of 8--26\,km\,s$^{-1}$
as shown by our observations.
Their rotation periods are significantly larger than those for old brown dwarfs
(below one day, e.g. Bailer-Jones \& Mundt 2001) 
and their rotational velocities are on average smaller 
than for old brown dwarfs (10--60\,km\,s$^{-1}$, Mohanty \& Basri 2003).
This is in agreement with the idea that they are in an early contracting stage
and will further spin up and contract.

Periodic photometric variabilities have been also detected for 
two dozen substellar $\sigma$\,Ori and $\epsilon$\,Ori members and member candidates
(Bailer-Jones \& Mundt 2001, Zapatero Osorio et al. 2003, Scholz \& Eisl\"offel
2004a,b, Caballero et al. 2004).
Including our periods in Cha\,I, the to-date known photometric periods for young brown 
dwarfs cover a wide range from 46\,min to 3.4 days.
Apart from rotational modulation due to chromospheric spots, several other possible
processes have been suggested to account for the periodic variabilities, 
like accretion phenomena, formation and evolution of dust coverage
for the cooler ones and eclipsing companions, which could all occur on various time scales.
Among all known periods for young brown dwarfs, 
only four have been confirmed as rotational periods by 
spectroscopic measurements of their rotational velocities $v \sin i$ 
(Cha\,H$\alpha$\,2, Cha\,H$\alpha$\,3, Cha\,H$\alpha$\,6, S~Ori\,25, 
Joergens \& Guenther 2001, Joergens et al. 2003b, Muzerolle et al. 2003,
Caballero et al. 2004), they all lie between 1.7 and 3.5 days.
Several periods of the order of a few hours have been also interpreted as rotational periods. 
However, as noticed by Joergens et al. (2003b), 
rotation periods of a few hours for very young brown dwarfs 
imply extreme rotation with rotational velocities of 100\,km\,s$^{-1}$ or more given the still 
large radii at this young age. These speeds come close to break-up velocities and 
for the 46\,min period, Zapatero Osorio et al. (2003) found that it 
is indeed above break-up velocity and cannot be a rotation period.
The finding of a rotation of 100\,km\,s$^{-1}$ or more
for brown dwarfs at a few million years would also be surprising 
in respect of the fact that they are in an early contracting stage and the 
expectation of a further spin-up in their future evolution.
A check of the proposed extreme rapid rotation by $v \sin i$ measurements would be
desirable. 
So far, all $v \sin i$ determinations for young brown dwarfs in Cha\,I, Taurus
and $\sigma$ Ori indicate rotational velocities below 26\,km\,s$^{-1}$ with the vast majority  
below 20\,km\,s$^{-1}$ (Joergens \& Guenther 2001, White \& Basri 2003, 
Muzzerolle et al. 2003).

\section{Summary and conclusions}

The presented work reports about observations of a population of 
very young brown dwarfs and low-mass stars close to the 
substellar borderline in the Cha\,I star forming region. 
At an age of only a few million years, their exploration
allows insights into the formation and early evolution of brown dwarfs.
The targets were studied in terms of their kinematic properties,
the occurrence of multiple systems among them as well as their rotational
characteristics. 

We are carrying out a RV survey for planets and brown dwarf companions to 
the targets with UVES at the VLT. The achieved RV precision 
allows us to search for companions down to Jupiter mass planets.
The analysis of the high-resolution spectra reveals very constant RVs on time scales 
of weeks to months for the majority of the targets as well as RV variability for three sources 
(Joergens et al. 2005a).
The RV constant objects are six brown dwarfs and three very low-mass stars 
(M$\leq$0.12\,M$_{\odot}$, spectral types M5--M8), for which we estimate 
upper limits for masses
of hypothetical companions to lie between 0.1\,M$_\mathrm{Jup}$ and 1.5\,M$_\mathrm{Jup}$.
This group shows a relation of decreasing RV amplitude with 
increasing mass. This reflects simply a higher RV precision for more massive targets
due to a better signal-to-noise, whereas
the effect of RV errors caused by surface features would be the opposite. 
This demonstrates
that brown dwarfs and very low-mass stars (M$\leq$0.12\,M$_{\odot}$) in Cha\,I display no
significant RV noise due to surface spots down to the precision necessary to detect 
Jupitermass planets. Thus, they are suitable targets to search 
for planets with the RV technique. 

Three objects exhibit significant RV variations with peak-to-peak amplitudes of 
2--3\,km\,s$^{-1}$:
the brown dwarf Cha\,H$\alpha$\,8 and the low-mass stars CHXR\,74 
($\sim$0.17\,M$_{\odot}$) and Sz\,23 ($\sim$0.3\,M$_{\odot}$).
For Sz\,23, which is the highest mass object in our sample, 
we have indications that the RV variations
are caused by surface spots from the variability time scale and from
significant CaII IR triplet emission.
Cha\,H$\alpha$\,8 and CHXR\,74 show a different variability behaviour with displaying
only very small amplitude or no variations on time scales of days to weeks but significant 
RV variations on times scales of months or longer, which cannot be explained by being 
rotational modulation. 
If caused by orbiting companions, the detected RV variations of CHXR\,74 and 
Cha\,H$\alpha$\,8 correspond to giant planets of a few Jupiter masses with periods of 
several months.
In order to explore the nature of the detected RV variations 
follow-up observations of CHXR\,74 and Cha\,H$\alpha$\,8 are planned. 
If confirmed as planetary systems, they would be unique because 
they would contain not only the lowest mass primaries 
and the first brown dwarf with a planet but with an age of a few million years
also the by far youngest extrasolar planets found to date. That would provide
empirical constraints for planet formation time scales as well as for the formation
of brown dwarfs.

The found multiplicity fraction in this survey is obviously very small.
Considering the subsample of the ten brown dwarfs and very low-mass stars with masses 
$\leq$ 0.12\,M$_{\odot}$,
only one of them (Cha\,H$\alpha$\,8) shows signs of RV variability, while the others
are RV quiet with respect to both companions and spots in our observations.
That hints at a very small multiplicity fraction of 10\% or less.
However, the RV variable brown dwarf Cha\,H$\alpha$\,8 and also the higher mass CHXR\,74
(not included in the above considered subsample)
are hinting at the possibility that companions to young brown dwarfs and very low-mass stars
have periods of several months and such a time scale was not 
covered for all targets. Thus, our results of small multiplicity reflects so far mainly 
separations of around 0.1\,AU. Further 3rd epoch RV determinations are planned.
At much larger separations, a direct imaging search for wide
(planetary or brown dwarf) companions to mostly the same targets 
also found a very small multiplicity fraction of $\leq$10\% (Neuh\"auser et al. 2002, 2003).
There still remains a significant gap in the studied separation ranges, which 
will partly be probed by the planned follow-up 3rd epoch RV measurements and partly is only 
accessible with high-resolving AO imaging (NACO\,/\,VLT) or even requires 
interferometric techniques (e.g. AMBER at the VLTI). 
For the already studied separations, the overall picture is a multiplicity 
fraction significantly smaller than for T~Tauri stars even when taking into account
the smaller available mass range for the companions. This hints at differences 
in the formation processes of brown dwarfs and T~Tauri stars.



In order to test the proposed ejection scenario for the formation of brown dwarfs,
we explored the kinematic properties of our substellar targets
based on absolute mean RVs derived within the presented RV survey.
We find that the brown dwarfs in Cha\,I form also kinematically a very homogeneous group.
They have very similar absolute RVs with a RV dispersion in terms of standard deviation of 
only 0.9\,km\,s$^{-1}$ (Joergens \& Guenther 2001, Joergens et al. 2005b).
A study of T~Tauri stars in the same field showed that there are no 
indications for a more violent dynamical evolution, like more frequent ejections,
for the brown dwarfs compared to the T~Tauri stars since
the RV dispersion of the T~Tauri stars (1.3\,km\,s$^{-1}$) was determined to be even 
slightly larger than that for the brown dwarfs.
This is the first observational constraint for the velocity distribution
of a homogeneous group of closely confined very young brown
dwarfs and therefore a first \emph{empirical} upper limit for ejection velocities.

Theoretical models of the ejection scenario have been performed by several groups in recent years.
Some of them are hinting at the possibility 
of a only small or of no mass dependence of the velocities.
Thus, the fact that we do not find a larger velocity dispersion for the brown dwarfs
than for the T~Tauri stars does not exclude the ejection scenario. 
However, we observe smaller velocities than any of the theoretical predictions for brown dwarfs formed
by the ejection scenario.
This might be partly attributed to the fact that our sample is statistically relatively small,
or it might be explained by 
the lower densities in Cha\,I compared to some model assumptions, or by shortcomings in the models, like 
neglection of feedback processes (Bate et al. 2003, Bate \& Bonnell 2005, Delgado-Donate et al. 2004) 
or of the cluster potential (Sterzik \& Durisen 2003, Umbreit et al. 2005).
The current conclusion is that either the brown dwarfs in 
Cha\,I have been formed by ejection but with smaller velocities as theoretically predicted 
or they have not been formed in that way.
We are planning to enlarge the sample in the future to put the results on an
improved statistical basis.


Finally, we studied the rotational properties of the targets in terms of 
projected rotational velocities $v\,\sin i$ measured in UVES spectra
as well as in terms of absolute rotational periods derived from light curve analysis.
We found that the $v\,\sin i$ values of the bona fide and candidate brown dwarfs in Cha\,I 
range between 8\,km\,s$^{-1}$ and 26\,km\,s$^{-1}$. These were the first determinations 
of rotational velocities for very young brown dwarfs (Joergens \& Guenther 2001).
Furthermore, we have determined rotational periods, consistent with $v \sin i$ values,
by tracing modulations of the light curves due to surface spots at the rotation period. 
We found periods for the three brown dwarf candidates
Cha\,H$\alpha$\,2, Cha\,H$\alpha$\,3 and Cha\,H$\alpha$\,6
of 2--3 days and for two very low-mass stars
B\,34 and CHXR78C (M$\leq$0.12\,M$_{\odot}$) of 4--5 days (Joergens et al. 2003b).
Magnetic braking due to 
interactions with a circum-stellar disk may play a role for some of them 
since the ones with detected disks are the slower rotators.

The emerging picture of the rotation of young brown dwarfs at an age of a 
few million years based on rotational velocities for brown dwarfs in Cha\,I,
Taurus and $\sigma$ Ori 
(Joergens \& Guenther 2001, White \& Basri 2003, Muzzerolle et al. 2003) 
and on spectroscopically confirmed rotation periods (our three
brown dwarfs in Cha\,I plus one $\sigma$\,Ori brown dwarf, Caballero et al. 2004)
indicates that young brown dwarfs
rotate with periods of the order of a few days and speeds of 7 to 26\,km\,s$^{-1}$.
Their rotation periods are significantly larger than those for old brown dwarfs
(below one day, e.g. Bailer-Jones \& Mundt 2001) 
and their rotational velocities are on average smaller 
than for old brown dwarfs (10--60\,km\,s$^{-1}$, e.g. Mohanty \& Basri 2003).
This is in agreement with the idea that they are in an early contracting stage
and will further spin up and contract before they reach a final radius when  
their interior electrons are completely degenerate. 

Despite the fact that the origins of brown dwarfs are still shrouded in mist, we think that the
presented comprehensive observations of very young brown dwarfs in Cha\,I
and the determination of their fundamental parameters
brought us an important step forward in revealing the details of one of the main
open issues in stellar astronomy and in the origins of solar systems.

\subsection*{Acknowledgements}
It is a pleasure to acknowledge fruitful collaborations
in the last years on the subject of this article
with Ralph Neuh\"auser, Eike Guenther,
Matilde Fern\'andez, Fernando Comer\'on and G\"unther Wuchterl.
Furthermore, I am grateful for a grant from the 
'Deutsche Forschungsgemeinschaft' 
(Schwerpunktprogramm `Physics of star formation') during my PhD time 
as well as current support by the European Union through a Marie Curie 
Fellowship under contract number FP6-501875.

\subsection*{References}

{\small

\bref
Allard F., Hauschildt P.H., Alexander D.R., Starrfield S. 1997 ARA\&A 35, 137

\bref
Armitage, P. J.; Clarke, C. J. 1997, MNRAS, 285, 540

\bref
Bailer-Jones C.A.L., Mundt R. 2001, A\&A 367, 218


\bref
Basri G., Mart\'{\i}n E.L. 1999, ApJ 118, 2460 

\bref
Basri G. 2000, ARA\&A 38, 485 



\bref
Bate M.R., Bonnell I.A., Bromm V. 2003, MNRAS 339, 577

\bref
Bate M.R., Bonnell I.A. 2005, MNRAS 356, 1201 

\bref
B\'ejar V.J.S., Zapatero Osorio M.R., Rebolo R. 1999, ApJ 521,
        671

\bref
Burrows A., Sharp C.M. 1999, ApJ 512, 843

\bref
Burrows A., Hubbard W.B., Lunine J.I., Liebert J. 2001, Rev. Mod. Phys. 73, 3

\bref
Caballero J.A., B\'ejar V.J.S., Rebolo R., Zapatero Osorio M.R. 2004, A\&A 424, 857

\bref
Carpenter J.M., Hillenbrand L.A., Skutskie M.F., Meyer M.R. 2002, AJ 124, 1001

\bref
Chauvin G., Lagrange A.-M., Dumas C. et al. 2004, A\&A 425, L29

\bref
Comer\'on F., Rieke G.H., Neuh\"auser R. 1999, A\&A 343, 477 

\bref
Comer\'on F., Neuh\"auser R., Kaas A.A. 2000, A\&A 359, 269

\bref
Covino E., Alcal\'a J.M, Allain S. et al. 1997, A\&A 328, 187

\bref
Dekker H., D'Odorico S., Kaufer A., Delabre B., Kotzlowski H. 2000, 
        In: SPIE Vol. 4008, p. 534, ed. by M.Iye, A.Moorwood

\bref
Delfosse X., Tinney C.G., Forveille T. et al. 1997, A\&A 327, L25

\bref
Delgado-Donate E.J., Clarke C.J., Bate M.R. 2003, MNRAS 342, 926

\bref
Delgado-Donate E.J., Clarke C.J., Bate M.R. 2004, MNRAS 347, 759

\bref
Dubath P., Reipurth B., Mayor M. 1996, A\&A 308, 107 

\bref
Durisen R.H., Sterzik M.F., Pickett B.K. 2001, A\&A 371, 952

\bref
Dworetsky M.M. 1983, MNRAS 203, 917


\bref
Ghez A.M., Neugebauer G., Matthews K. 1993, AJ 106, 2005

\bref
Guenther E.W., Wuchterl G. 2003, A\&A 401, 677

\bref
Hawley S.L., Covey K.R., Knapp G.R. et al. 2002, AJ 123, 3409

\bref
Joergens V., Guenther E. 2001, A\&A 379, L9

\bref
Joergens V. 2003,
     PhD thesis, Ludwigs-Maximilians Universit\"at M\"unchen

\bref
Joergens\,V., Fern\'andez\,M., Carpenter\,J.M.,\,Neuh\"auser\,R. 2003b,\,ApJ\,594, 971

\bref Joergens V., Neuh\"auser R., Guenther E.W., Fern\'andez M., Comer\'on F.,
    In: IAU Symposium No. 211, Brown Dwarfs,
     ed. by E. L. Mart\'\i n, Astronomical Society of the Pacific, San Francisco,
     2003a, p.233

\bref
Joergens V. et al. 2005a, in prep. 

\bref 
Joergens V. et al. 2005b, in prep. 

\bref
Kirkpatrick J.D., Reid I.N., Liebert J. et al. 2000, AJ 120, 447

\bref
K\"ohler R., Kunkel M., Leinert Ch., Zinnecker H. 2000, A\&A 356, 541

\bref
Kroupa, P., Bouvier, J. 2003, MNRAS 346, 369

\bref
Kumar S. 1962, AJ 67, 579

\bref
Kumar S. 1963, ApJ 137, 1121 	

\bref
Leinert Ch., Zinnecker H., Weitzel N. et al. 1993, A\&A 278, 129


\bref
Low C., Lynden-Bell D. 1976, MNRAS 176, 367

\bref
Luhman K. 2004, ApJ 602, 816


\bref
Mart\'\i n E.L., Brandner W., Bouvier J. et al. 2000 ApJ 543, 299

\bref
Mart\'\i n, E. L., Zapatero Osorio M. R. \& Lehto H.J. 2001, ApJ, 557, 822

\bref
Mayor M., Queloz D. 1995, Nature 378, 355

\bref
Mayor, M., Udry, S., Naef, D. et al. 2003, A\&A 415, 391

\bref
Mizuno A., Hayakawa T., Tachihara K. et al. 1999, PASJ 51, 859

\bref
Mohanty S., Basri G. 2003, ApJ 583, 451

\bref
Muzerolle J., Hillenbrand L., Calvet N., Brice\~no C., Hartmann L. 2003, ApJ, 592, 266

\bref
Nakajima T., Oppenheimer B.R., Kulkarni S.R. et al. 1995, Nature 378, 463

\bref
Neuh\"auser R., Comer\'on F. 1998, Science 282, 83 

\bref
Neuh\"auser R., Comer\'on F. 1999, A\&A 350, 612

\bref
Neuh\"auser R., Brandner W., Alves J., Joergens V., Comer\'on F. 2002, 
		A\&A 384, 999

\bref
Neuh\"auser R., Brandner W., Guenther E. 2003, IAUS 211, 309, ed. by E. Mart\'\i n

\bref
Oppenheimer B.R., Kulkarni S.R., Matthews K., Nakajima T. 1995
        Science 270, 1478

\bref
Rebolo R., Mart\'\i n E.L., Magazz\'u A. 1992, ApJ 389, L83

\bref
Rebolo R., Zapatero Osorio M.R., Mart\'\i n E.L. 1995, Nature 377, 
        129

\bref
Reipurth B., Clarke C. 2001, ApJ 122, 432


\bref
Santos N.C., Bouchy F., Mayor M., Pepe F., Queloz D. et al. 2004, A\&A 426, L19

\bref
Scholz A., Eisl\"offel J. 2004a, A\&A 419, 249

\bref
Scholz A., Eisl\"offel J. 2004b, A\&A in press, astro-ph/0410101

\bref
Shu F.H., Adams F.C., Lizano S. 1987, ARA\&A 25, 23    

\bref
Stauffer J.R., Hamilton D., Probst R. 1994, AJ 108, 155 

\bref
Sterzik M.F., Durisen R.H. 1995, A\&A 304, L9

\bref
Sterzik M.F., Durisen R.H. 1998, A\&A 339, 95

\bref
Sterzik M.F., Durisen R.H. 2003, A\&A, 400, 1031


\bref
Tsuji T., Ohnaka K., Aoki W. 1996a, A\&A 305, L1

\bref
Tsuji T., Ohnaka K., Aoki W., Nakajima T. 1996b, A\&A 308, L29

\bref
Umbreit S., Burkert A., Henning T., Mikkola S., Spurzem R. 2005, A\&A, in press

\bref
Valtonen M., Mikkola S. 1991, ARA\&A 29, 9

\bref
Walter F.M. 1992, AJ 104, 758

\bref
White, R. J. \& Basri, G. 2003, ApJ 582, 1109

\bref
Whitworth, A.P., Zinnecker, H. 2004, A\&A 427, 299

\bref

\bref
Wuchterl G., Guillot T., Lissauer J.J. 2000.
        In: \emph{Protostars and Planets IV}, 
	ed. by Mannings V., Boss A.P., Russell S.S., 
	Univ. of Arizona Press, Tucson, p.1081

\bref
Zapatero Osorio M.R., Caballero J.A., B\'ejar V.J.S., Rebolo R. 2003, A\&A 408, 663

}

\vfill

\end{document}